\documentclass[journal,letter,10pt,twocolumn]{IEEEtran}
\pdfoutput=1

\usepackage{caption}
\usepackage{amsmath, amssymb, amscd, amsthm, amsfonts}
\usepackage{bbm}
\usepackage[pdftex]{graphicx}
\usepackage[caption=false,font=footnotesize]{subfig}
\usepackage{dblfloatfix}    % To enable figures at the bottom of page
\usepackage{color}
\usepackage{algorithm}
\usepackage[noend]{algpseudocode}
\usepackage{comment}

\usepackage{hyperref}

\usepackage{tikz,ifthen}
\usetikzlibrary{positioning,arrows,backgrounds}
\usetikzlibrary{fit}
\usetikzlibrary{calc}
 \usepackage{color}
 \usepackage{transparent}

\newtheorem{theorem}{Theorem}

\newtheorem{corollary}{Corollary}
\newtheorem{proposition}{Proposition}
\newtheorem{definition}{Definition}
\newtheorem{assumption}{Restriction}

\usepackage{pgfplots}
\pgfplotsset{width=10cm,compat=1.9}

\usepackage{subfiles}
\usepackage[noadjust]{cite}

\usepackage[english]{babel}

\addto\captionsenglish{}

\graphicspath{{../Figures/}}



\begin{document}

\title{Design and Analysis of Hardware-limited Non-uniform Task-based Quantizers}

\author{
   \IEEEauthorblockN{Neil Irwin Bernardo, \textit{Graduate Student Member, IEEE}, Jingge Zhu, \textit{Member, IEEE}, Yonina C. Eldar, \textit{Fellow, IEEE}, and Jamie Evans, \textit{Senior Member, IEEE}
   }
   \thanks{\textcolor{black}{Part of this work has been accepted for presentation in IEEE International Conference on Acoustics, Speech, and Signal Processing 2023 (ICASSP 2023). The codes of this paper are available in the following Code Ocean capsule: \url{https://codeocean.com/capsule/0782084/tree/v1}}}
    \thanks{\textcolor{black}{This project has received funding from the Australian Research Council under project DE210101497, from the European Research Council (ERC) under the European Union’s Horizon 2020 research and innovation programme (grant agreement No. 101000967), and from the Israel Science Foundation (grant No. 536/22). N.I. Bernardo acknowledges the Melbourne Research Scholarship of the University of Melbourne and the DOST-ERDT Faculty Development Fund of the Republic of the Philippines for sponsoring his doctoral studies.}}
   
   \thanks{N.I. Bernardo is with the Department of Electrical and Electronic Engineering, The University of Melbourne, Parkville, VIC 3010, Australia and also with the Electrical and Electronics Engineering Institute, University of the Philippines Diliman, Quezon City 1101, Philippines (e-mail: bernardon@student.unimelb.edu.au).}
    \thanks{J. Zhu and J. Evans are with the Department of Electrical and Electronic Engineering, The University of Melbourne, Parkville, VIC 3010, Australia (e-mail: jingge.zhu@unimelb.edu.au;
jse@unimelb.edu.au).}
      \thanks{Y. C. Eldar is with the Faculty of Math and CS, Weizmann Institute of Science, Rehovot 7610001, Israel (e-mail: yonina.eldar@weizmann.ac.il).}
}

\maketitle
\begin{abstract}
Hardware-limited task-based quantization is a new design paradigm for data acquisition systems equipped with serial scalar analog-to-digital converters using a small number of bits. By taking into account the underlying system task, task-based quantizers can efficiently recover the desired parameters from the low-bit quantized observation. Current design and analysis frameworks for hardware-limited task-based quantization are only applicable to inputs with bounded support and uniform quantizers with non-subtractive dithering. Here, we propose a new framework based on generalized Bussgang decomposition that enables the design and analysis of hardware-limited task-based quantizers that are equipped with non-uniform scalar quantizers or that have inputs with unbounded support. We first consider the scenario in which the task is linear. Under this scenario, we derive new pre-quantization and post-quantization linear mappings for task-based quantizers with mean squared error (MSE) that closely matches the theoretical MSE. Next, we extend the proposed analysis framework to quadratic tasks. We demonstrate that our derived analytical expression for the MSE accurately predicts the performance of task-based quantizers with quadratic tasks.
\end{abstract}


\IEEEpeerreviewmaketitle

\section{Introduction}\label{section-intro}

\IEEEPARstart{D}{igital} systems are equipped with quantizers to facilitate the processing, storage, and communication of information embedded in continuous-amplitude samples. In principle, the most accurate digital representation of a sampled signal is obtained by jointly mapping the samples to the digital domain via vector quantization \cite{Gersho:1991,Gray:1998}. The optimal trade-off between compression and fidelity is fundamentally described by rate-distortion theory \cite{Berger:1998}. However, in practice, the quantization process is performed by analog-to-digital converters (ADC) which typically operate in a serial scalar manner \cite{Eldar:2015}. Under this setup, the incoming continuous-time analog signal is first sampled and the samples are sequentially mapped by the quantizer in digital form using a finite number of quantization bits \cite{Kosonocky:1999}. A linear increase in the number of quantization bits corresponds to an exponential increase in power consumption \cite{Walden:1999}. Therefore, there is growing interest in the use of low-resolution data converters. For instance, recent works on low-power communication receivers, such as \cite{bernardo2021sep,bernardo2021TIT,bernardo2022polar,Choi:2020,Wen:2015,Schluter:2020}, have focused on investigating the performance limits of low-resolution receiver architectures and designing novel methods that enable various receiver functionalities (detection, channel estimation, and synchronization) to work in the low-resolution regime.

\begin{figure*}[t]
    \centering
    \includegraphics[scale = .725]{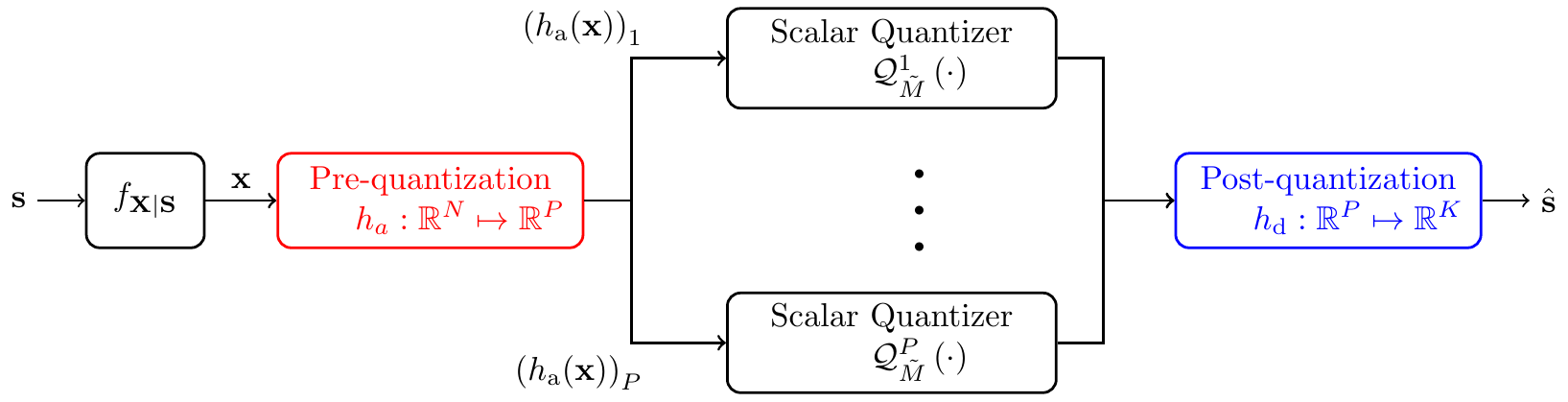}
    \captionsetup{justification=centering}
    \caption{System model for task-based quantization.}
    \label{fig:tb_general}
\end{figure*}

Data acquisition systems are often designed such that the input and output of the quantizers are close with respect to some pre-defined distortion measure \cite[Ch. 10]{Cover:2006_IT}\cite{Kipnis2:2018}. This design approach, however, does not take into account the underlying system task. In several signal processing and communication applications, the objective is not to faithfully recover the input signal, but rather to extract some low-dimensional parameters/features embedded in the quantized measurements. Such systems that take into account the underlying task are generally referred to as \emph{task-based quantization}, and task-based quantization systems equipped with serial scalar ADCs are specifically referred to as \emph{hardware-limited task-based quantization} \cite{Shlezinger:2019}. 

Previous works \cite{Shlezinger:2019,Shlezinger2:2019,Shlezinger:2020,Salamatian:2019} have shown that, by exploiting the \emph{a priori} knowledge regarding the system task, hardware-limited task-based quantizers can outperform digital systems that simply extract the desired parameters from the quantized measurements. Performance gain in task-based quantization is achieved by employing a hybrid analog/digital (A/D) architecture and jointly designing the analog pre-quantization mapping and digital post-quantization mapping in light of the underlying system task. The task-based quantization framework has been applied in various tasks such as channel estimation \cite{Shlezinger:2019}, empirical covariance estimation \cite{Salamatian:2019}, multiple-input multiple-output (MIMO) radar receivers \cite{Xi:2021}, task-specific beamforming \cite{Zirtiloglu:2022}, MIMO communication \cite{Rini:2017,Khalili:2021,bernardo2022MIMO}, symbol detection \cite{Khobahi:2021,Khobahi2:2021}, and graph signal compression \cite{Li:2021}. Moreover, the combined effect of sampling and quantization in hardware-limited task-based systems has been studied in \cite{Neuhaus:2021,Neuhaus2:2021} and it is shown that the optimal performance of bitrate-constrained data acquisition systems is generally achieved by sampling below the Nyquist rate. The optimal sampling and quantization scheme for task-based data acquisition may also be learned via data-driven approaches if the input distribution is not known \cite{Shlezinger:2021}.

Despite the aforementioned benefits and the wide range of applications of task-based quantization, the existing framework for analyzing hardware-limited task-based quantization is only applicable to scalar uniform ADCs with non-subtractive dithering. While dithering offers analytical tractability, dithered quantizers generally have subpar performance compared to their non-dithered counterpart \textcolor{black}{when the input has a bandlimited characteristic function because of their increased quantization noise energy  \cite{Shlezinger:2019,Widrow:1996}}. As such, the current framework does not fully capture the actual performance of task-based quantization. Simulation results of \cite{Shlezinger:2019} depict large performance gaps between the dithered and non-dithered case when the number of quantization levels per scalar quantizer is low. Also, the theory of nonsubtractive dithering \cite{Wannamaker:2000} only applies to uniform quantizers. Mathematical tools \cite{Akyol:2013} for analyzing dithered non-uniform quantization exist but only for subtractive dithering. Furthermore, the analysis framework relies on the assumption that the overload probability (i.e. the probability that the ADC input does not exceed the specified dynamic range of the ADC) is zero. This assumption can be quite restrictive so that the analysis framework only holds approximately for input signals with infinite support. Guidelines on how to set the overload probability for a given number of quantization levels are provided in \cite{Neuhaus:2021}. Still, the simulated distortion of dithered task-based quantization is approximately 5\% higher than what the analytical expression predicts in the numerical results.

Here, we provide a new approach to design and analyze hardware-limited task-based quantization systems with analog pre-quantization and digital post-quantization linear mappings based on generalized Bussgang decomposition \cite{Demir:2021}. In contrast to the state-of-the-art (SOTA) analysis framework \cite{Shlezinger:2019}, the proposed framework does not rely on the zero overload probability assumption and is also applicable to non-uniform scalar quantizers and non-dithered settings. Our proposed framework restricts the pre-quantization mapping to be within the class of linear mappings that make the inputs of the scalar quantizers uncorrelated. While this restriction may lead to suboptimal performance, our numerical results show that the proposed framework can achieve lower distortion than previous results when the quantization budget is limited. More importantly, a crucial advantage of our analysis is that the simulated distortion of task-based quantizers designed using our method fits well with the predictions of our theoretical framework, even if the underlying system task is nonlinear. This is in contrast to previous results which only hold approximately. The main contributions of our work are the following:
\begin{itemize}
    \item We provide descriptions of the analog and digital linear mappings of task-based quantization under a linear task assumption (i.e. the task is a linear function of the observations). The derived linear mappings are conceptually different from the linear mappings in previous works. We present numerical results showing that, in some cases, task-based quantizers designed using our approach can outperform task-based quantizers designed using the SOTA analysis framework \cite{Shlezinger:2019}.
    \item We show that the actual mean squared error (MSE) of the task-based system under the derived analog and digital linear mappings fits the theoretical MSE in contrast to previous results. Moreover, the proposed analysis framework also enables a model-based analysis of task-based quantization with non-uniform quantizers. To the best of our knowledge, there is no framework in the literature that facilitates model-based analysis of task-based quantization with non-uniform quantizers.
    %The proposed analysis framework also enables a model-based analysis of task-based quantization with non-uniform quantizers. To the best of our knowledge, there is no framework in the literature that facilitates model-based analysis of task-based quantization with non-uniform quantizers.
    %\item We show that the actual mean squared error (MSE) of the task-based system under the derived analog and digital linear mappings fits the theoretical MSE. Moreover, when non-uniform quantizers are used, the MSE of the task-based quantization system under the proposed analysis framework is lower than the simulated MSE of SOTA analysis framework \cite{Shlezinger:2019} for linear tasks.
    \item We show how to extend the proposed framework to nonlinear tasks. More specifically, we consider the quadratic task problem of empirical covariance estimation and show that the task-based quantization system designed using our proposed framework achieves lower MSE than the simulated MSE of the task-based system designed using the framework presented in \cite{Salamatian:2019}.  
\end{itemize}

The rest of the paper is organized as follows: Section \ref{section-problem_statement} formulates the system model and states the model assumptions for the linear task scenario. Section \ref{section-main_results} presents the new analysis framework. Section \ref{section-numerical_linear} provides numerical results and analysis for the proposed framework in Section \ref{section-main_results}. Section \ref{section-quadtask} extends the developed framework to quadratic tasks. Finally, Section \ref{section-conclusion} concludes the paper.

\section{Problem Formulation and Analysis Tools}\label{section-problem_statement}

\subsection{Problem Setup and Model Assumptions}\label{section-problem_setup}

The system model of the task-based quantization with hardware constraints is illustrated in Figure \ref{fig:tb_general}. The \emph{task vector} $\mathbf{s} \in \mathbb{R}^{K\times 1}$ contains the parameters we aim to recover. However, the input to the task-based quantizer is not $\mathbf{s}$ but the \emph{measurement vector} $\mathbf{x}\in\mathbb{R}^{N\times 1}$. The statistical relationship between $\mathbf{s}$ and $\mathbf{x}$ is described by the conditional probability $f_{\mathbf{X}|\mathbf{S}}(\mathbf{x}|\mathbf{s})$. With slight abuse of notation, we simply write the conditional probability as $f_{\mathbf{X}|\mathbf{S}}$. We also assume that $\mathbf{s}$ and $\mathbf{x}$ are both zero-mean random vectors and have covariance matrices given by $\boldsymbol{\Sigma}_{\mathbf{s}}$ and $\boldsymbol{\Sigma}_{\mathbf{x}}$, respectively.

The measurement vector $\mathbf{x}$ is projected to $\mathbb{R}^{P\times 1}$, where $P \leq N$, using an analog pre-quantization mapping, denoted $h_{\mathrm{a}}$. The $P$ outputs of the analog pre-quantization mapping are fed to $P$ scalar quantizers. From \cite[Corollary 1]{Shlezinger:2019}, the optimal choice of $P$ must not exceed $K$. Each scalar quantizer has $\tilde{M} = \lfloor M^{\frac{1}{P}}\rfloor$ number of quantization levels, where $M$ is a constraint on the overall number of quantization levels. As pointed out in \cite{Shlezinger:2019}, the parameter $\tilde{M}$ is directly related to the power consumption of an ADC. We allow the quantization levels of the scalar quantizers to have non-uniform and non-identical structure. In this work, we assume that the scalar quantizers are designed using the Lloyd-Max algorithm \cite{Lloyd:1982}. Note that quantizers designed using the Lloyd-Max algorithm satisfy $\mathbb{E}\{X_{\mathrm{in}}|\mathcal{Q}(X_{\mathrm{in}})\} = \mathcal{Q}(X_{\mathrm{in}})$, where $X_{\mathrm{in}}$ is the input to the quantizer (i.e. \textcolor{black}{the representative level of a quantization interval is its conditional mean value}). This property of the scalar quantizers is crucial in the derivation of our main results. \textcolor{black}{Note, however, that Lloyd-Max algorithm does not necessarily produce the globally-optimal quantizer, unless the quantizer input has a log-concave distribution \cite{Kieffer:1983}.} Finally, the $P$ outputs of the scalar quantizers, denoted $\mathbf{z}\in\mathbb{R}^{P\times 1}$, are fed to a digital post-processing function to estimate the task vector $\mathbf{s}$. We represent the estimate of the task vector as $\mathbf{\hat{s}}\in \mathbb{R}^{K\times 1}$.

%%Note, however, that the analysis framework we present here holds even for uniform scalar ADCs provided that the quantizer design satisfies $\mathbf{E}\{X_{\mathrm{in}}|\mathcal{Q}(X_{\mathrm{in}})\} = \mathcal{Q}(X_{\mathrm{in}})$, where $X_{\mathrm{in}}$ is the input to the quantizer (i.e. each quantization interval is represented by its mean value). By design, uniform and non-uniform scalar quantizers that are constructed using Lloyd-Max algorithm satisfy this property. 

\begin{figure*}[t]
    \centering
    \includegraphics[scale = .725]{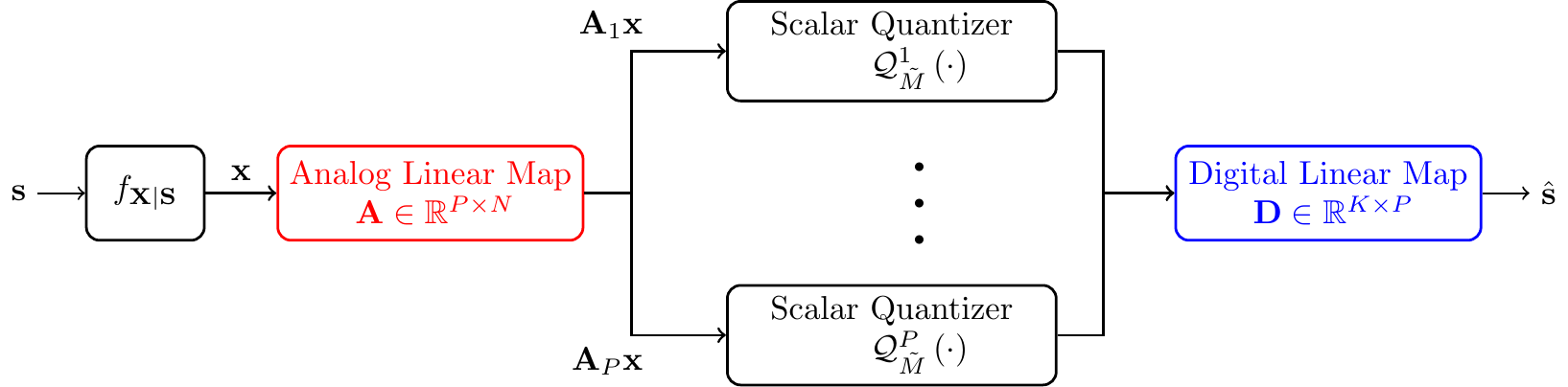}
    \captionsetup{justification=centering}
    \caption{Task-based quantization with analog combining matrix and digital processing matrix.}
    \label{fig:tb_linear}
\end{figure*}

The goal is to recover $\mathbf{s}$ from quantized measurements $\mathbf{z}$. \textcolor{black}{The general} problem setup is referred in the information theory community as the \emph{indirect source coding problem}\footnote{Other names used in the literature are remote source coding and noisy source coding.}\textcolor{black}{/\emph{indirect quantization}}, and was first introduced in \cite{Dobrushin:1962} \textcolor{black}{and \cite{Fine:1965}. The problem setup in this paper is different from the aforementioned works since the hardware-limited task-based quantizer structure is restricted to scalar ADCs.} We design $h_{\mathrm{a}}$ and $h_{\mathrm{d}}$ such that the MSE between $\mathbf{s}$ and $\mathbf{\hat{s}}$ is minimized. Mathematically, we have the following optimization problem:
\begin{align}\label{eq:MSE_objectivefunc}
    &\min_{h_{\mathrm{a}},h_{\mathrm{d}}}\mathbb{E}\left\{||\mathbf{s} - \mathbf{\hat{s}}||^2\right\}\nonumber \\
    &\quad\qquad\qquad= \mathbb{E}\left\{||\mathbf{s} - \mathbf{\tilde{s}}||^2\right\} + \min_{h_{\mathrm{a}},h_{\mathrm{d}}}\mathbb{E}\left\{||\mathbf{\tilde{s}} - \mathbf{\hat{s}}||^2\right\},
\end{align}
where $\mathbf{\tilde{s}} = \mathbb{E}\left\{\mathbf{s}|\mathbf{x}\right\}$ is the minimum MSE (MMSE) estimator of $\mathbf{s}$ given the measurement vector $\mathbf{x}$. The RHS of \eqref{eq:MSE_objectivefunc} shows that the MSE can be written as a sum of two terms. The first term quantifies the minimum estimation error of $\mathbf{s}$ from $\mathbf{x}$ whereas the second term accounts for the minimal distortion in quantizing the MMSE estimate. The first term is independent on the actual structure of the scalar quantizers and design of the pre- and post-quantization mappings \cite{Shlezinger:2019}. Thus, we can focus our attention on minimizing the second MSE term. We shall refer to this MSE term as the \emph{quantizer-dependent MSE}.

To facilitate recovery of the task vector $\mathbf{s}$ under practical hardware setting, we follow the approach of \cite{Shlezinger:2019} which is to impose $h_{\mathrm{a}}$ and $h_{\mathrm{d}}$ to be linear mappings, as shown in Figure \ref{fig:tb_linear}. That is, we introduce an analog combining matrix $\mathbf{A}\in\mathbb{R}^{P\times N}$ and a digital processing matrix $\mathbf{D}\in\mathbb{R}^{K\times P}$ to operate as $h_{\mathrm{a}}$ and $h_{\mathrm{d}}$, respectively. The quantities $\{\mathbf{A}_{p}\mathbf{x}\}_{p = 1}^{P}$ are the $P$ outputs of the analog combining matrix. The use of linear mappings in the analog and digital domain offers a lot of benefits from a practical viewpoint and is already done in various hybrid A/D receiver architectures (see \cite{Mo:2017,Roth:2017,Song:2016,Ioushua:2019}). 

Similar to previous works, we further relax the problem by considering linear tasks, i.e. $\mathbf{\tilde{s}} = \boldsymbol{\Gamma}\mathbf{x}$ for some $\boldsymbol{\Gamma}\in\mathbb{R}^{K\times N}$. Under the linear task scenario, we are able to derive closed-form expressions for $\mathbf{A}$, $\mathbf{D}$, and MSE than what were obtained in the previous work. The framework developed for the linear task will be later extended to the more general nonlinear tasks.

\subsection{Analysis Techniques for Task-based Quantization}

The previous work\cite{Shlezinger:2019} carried out the analysis assuming the system is equipped with non-subtractive uniform dithered quantizers. Whenever the input falls inside the dynamic range of a uniform dithered quantizer, the output can be written as a sum of the input and an additive zero-mean white quantization noise that is uncorrelated with the input. This simplication enables the derivation of the optimal linear mappings and MSE under a uniform dithered setting. Numerical results show that using these linear mappings on uniform undithered quantizers can further reduce the distortion. However, the theoretical framework established in \cite{Shlezinger:2019} is unable to accurately predict the actual MSE of the task-based quantizer. We present numerical results in Sections \ref{section-numerical_linear} and \ref{section-quadtask} to demonstrate this issue. Moreover, the non-subtractive dithering framework \cite{Wannamaker:2000} does not apply to non-uniform quantization.

To avoid the shortcomings of the previous work, we consider a different analysis technique to represent the output of the scalar quantizers in a more analytically tractable form. More precisely, we use the generalized Bussgang decomposition\footnote{\textcolor{black}{The original statement of the Bussgang Theorem \cite{Bussgang:1952} only applies to Gaussian signals. Instead, we use the generalized Bussgang decomposition mentioned in \cite[Section V-C]{Demir:2021} which works for non-Gaussian inputs.}} to represent the quantization process as a noisy linear function of the input. That is,
\begin{align}
    \mathbf{z} =& \mathcal{Q}_{\tilde{M}}^{1:P}(\mathbf{A}\mathbf{x})\nonumber\\
    =& \mathbf{B}\mathbf{A}\mathbf{x} + \boldsymbol{\eta}.
\end{align}
Here, $\mathcal{Q}_{\tilde{M}}^{i:j}(\cdot)$ denotes the outputs of the scalar quantizers from index $i$ to index $j$. The square matrix $\mathbf{B}\in\mathbb{R}^{P\times P}$ is called the Bussgang gain and $\boldsymbol{\eta} \in \mathbb{R}^{P\times 1}$ is the distortion vector uncorrelated with the quantizer input $\mathbf{A}\mathbf{x}$. The Bussgang gain matrix can be written as
\begin{align}\label{eq:Bussgang_gain}
    \mathbf{B} = \boldsymbol{\Sigma}_{\mathbf{zx}}\mathbf{A}^T\left(\mathbf{A}\boldsymbol{\Sigma}_{\mathbf{x}}\mathbf{A}^T\right)^{-1},
\end{align}
where $\boldsymbol{\Sigma}_{\mathbf{zx}}$ is the cross-covariance between $\mathbf{z}$ and $\mathbf{x}$. The covariance of the distortion vector, denoted $\boldsymbol{\Sigma}_{\boldsymbol{\eta}}$, can be expressed as
\begin{align}
    \boldsymbol{\Sigma}_{\boldsymbol{\eta}} = \boldsymbol{\Sigma}_{\mathbf{z}} - \mathbf{B}\left(\mathbf{A}\boldsymbol{\Sigma}_{\mathbf{x}}\mathbf{A}^T\right)\mathbf{B}^T,
\end{align}
where $\boldsymbol{\Sigma}_{\mathbf{z}}$ is the covariance of $\mathbf{z}$. 

The generalized Bussgang decomposition is exact; the intuition is that $\mathbf{B}\mathbf{A}\mathbf{x}$ is the linear MMSE estimate of $\mathbf{z}$ given the observation $\mathbf{A}\mathbf{x}$ (not necessarily Gaussian) \cite{Demir:2021}. However, the distribution of $\boldsymbol{\eta}$ is not known and the Bussgang gain matrix is, in general, not diagonal. Therefore, we introduce a restriction on the structure of $\mathbf{A}$ that makes the Bussgang gain matrix $\mathbf{B}$ diagonal, regardless of the distribution of the measurement vector $\mathbf{x}$.
%We then \textcolor{black}{provide a description of} the diagonal entries of $\mathbf{B}$.
\begin{assumption}\label{assumption:analogcombiner}
Suppose we denote $\mathbf{A}_{i}$ to be the $i$-th row of the analog combining matrix $\mathbf{A}$. Then, we pick $\mathbf{A}_{i}$ such that, for any $i\neq j$, we have $\mathbf{A}_{i}\boldsymbol{\Sigma}_{\mathbf{x}}^{\frac{1}{2}} \perp \mathbf{A}_{j}\boldsymbol{\Sigma}_{\mathbf{x}}^{\frac{1}{2}}$, where $\boldsymbol{\Sigma}_{\mathbf{x}}^{\frac{1}{2}}$ is the matrix square root of $\boldsymbol{\Sigma}_{\mathbf{x}}$.
\end{assumption}
Note that imposing Restriction \ref{assumption:analogcombiner} may yield sub-optimal task-based quantizer designs. In fact, we demonstrate in Section \ref{section-numerical_linear} that our design framework, which is based on Restriction \ref{assumption:analogcombiner}, does not always produce the task-based quantizer design with the lowest MSE. Nonetheless, our framework can achieve better performance than the current design and analysis frameworks when the scalar quantizers have very low resolution. Restriction \ref{assumption:analogcombiner} also forces the elements of $\boldsymbol{\eta}$ to be uncorrelated\footnote{The requirements mentioned in \cite[page 3]{Bjornson:2019} to make $\boldsymbol{\Sigma}_{\boldsymbol{\eta}}$ diagonal are satisfied since $\mathbf{A}\boldsymbol{\Sigma}_{\mathbf{x}}\mathbf{A}^T$ and $\mathbf{B}$ are diagonal.}. 
%Restriction \ref{assumption:analogcombiner} is reasonable since it forces the elements of $\boldsymbol{\eta}$ to be uncorrelated\footnote{The requirements mentioned in \cite[page 3]{Bjornson:2019} to make $\boldsymbol{\Sigma}_{\boldsymbol{\eta}}$ diagonal are satisfied since $\mathbf{A}\boldsymbol{\Sigma}_{\mathbf{x}}\mathbf{A}^T$ and $\mathbf{B}$ are diagonal.}.}

In the next section, we will show how imposing Restriction~\ref{assumption:analogcombiner} forces $\mathbf{B}$ to be diagonal. We will then use Restriction \ref{assumption:analogcombiner} in conjunction with the generalized Bussgang decomposition to establish a new framework for analyzing and designing hardware-limited task-based quantizers.

\section{Main Results}\label{section-main_results}

We now characterize the hardware-limited task-based quantizer which minimizes \eqref{eq:MSE_objectivefunc} under Restriction \ref{assumption:analogcombiner}. We first define a quantity that is crucial in stating the main results of the paper.
\begin{definition}\label{definition:rho_q}
Suppose $\mathbf{A}_{p}\in \mathbb{R}^{1\times N}$ is the $p$-th row of $\mathbf{A}$. Then, the distortion factor of the $p$-th quantizer, denoted $\rho_{\mathrm{q}}^{(p)}$, accounts for the relative amount of distortion introduced by the $p$-th quantizer to its input and is expressed as
\begin{align}
    \rho_{\mathrm{q}}^{(p)} = \frac{\mathbb{E}\{(z_{p} - \mathbf{A}_{p}\mathbf{x})^2\}}{\mathbb{E}\{(\mathbf{A}_{p}\mathbf{x})^2\}},
\end{align}
where $z_{p}$ is the output of the $p$-th scalar quantizer and the denominator term is the energy of the quantizer input.
\end{definition}
A typical scenario in which the MMSE estimator $\mathbf{\tilde{s}}$ is a linear function of $\mathbf{x}$ is when the task vector $\mathbf{s}$ and the measurement vector $\mathbf{x}$ are jointly Gaussian \cite[Section 3.2.7]{Hayes:1996}. Conveniently, the quantizer inputs $\mathbf{A}\mathbf{x}$ are also Gaussian. The distortion factor $\rho_{\mathrm{q}}^{(p)}$ for a Gaussian input and Lloyd-Max scalar quantizer is tabulated in \cite{Max:1960} for $\tilde{M} = 1$ up to $\tilde{M} = 36$ levels. For high-rate quantizers, the distortion factors of non-uniform and uniform quantizers under a Gaussian input are $\rho_{\mathrm{q}}^{(p)} \approx \frac{\pi\sqrt{3}}{2}\cdot \tilde{M}^{-2}$ and $\rho_{\mathrm{q}}^{(p)} \approx 1.47\cdot \tilde{M}^{-1.74}$, respectively \cite{Gray:1998,Max:1960}. 

The following proposition characterizes the diagonal entries of the Bussgang gain matrix under Restriction \ref{assumption:analogcombiner}.

\begin{proposition}\label{proposition:B_diagonal}
Under Restriction \ref{assumption:analogcombiner}, the Bussgang gain matrix $\mathbf{B}$ is a diagonal matrix and can be expressed as
\begin{align}
    \mathbf{B} = \mathrm{diag}\{\mathbf{1} - \boldsymbol{\rho}_{\mathrm{q}}\},
\end{align}
where the $\mathrm{diag}\{\cdot\}$ operator generates a $P\times P$ diagonal matrix with entries coming from the $P\times 1$ vector $\{\cdot\}$, and $\boldsymbol{\rho}_{\mathrm{q}} = [\rho_{\mathrm{q}}^{(1)},\rho_{\mathrm{q}}^{(2)},\cdots,\rho_{\mathrm{q}}^{(P)}]^T$.
\end{proposition}
\begin{proof}
See Appendix \ref{proof:B_diagonal}.
\end{proof}

We now present the main results of our work.
\begin{proposition}\label{proposition:D_opt}
For any analog combining matrix $\mathbf{A}$ that satisfies Restriction \ref{assumption:analogcombiner}, the optimal digital processing matrix, denoted $\mathbf{D}^\circ$, which minimizes the MSE is given by
\begin{align}
    \mathbf{D}^\circ\left(\mathbf{A}\right) = \boldsymbol{\Gamma}\boldsymbol{\Sigma}_{\mathbf{x}}\mathbf{A}^T(\mathbf{A}\boldsymbol{\Sigma}_{\mathbf{x}}\mathbf{A}^T)^{-1}.
\end{align}
Consequently, the quantizer-dependent MSE can be expressed as
\begin{align}
    &\mathbb{E}\{||\mathbf{\tilde{s}} - \mathbf{\hat{s}}||^2\}\nonumber\\
    &\qquad = \mathrm{Tr}\left(\boldsymbol{\Gamma}\boldsymbol{\Sigma}_{\mathbf{x}}\boldsymbol{\Gamma}^T\right)\nonumber \\
    &\qquad\quad- \mathrm{Tr}\left(\boldsymbol{\Gamma}\boldsymbol{\Sigma}_{\mathbf{x}}\mathbf{A}^T\mathbf{B}\left(\mathbf{A}\boldsymbol{\Sigma}_{\mathbf{x}}\mathbf{A}^T\right)^{-1}\mathbf{A}\boldsymbol{\Sigma}_{\mathbf{x}}\boldsymbol{\Gamma}^T\right)
\end{align}
\end{proposition}
\begin{proof}
See Appendix \ref{proof:D_opt}.
\end{proof}

\begin{theorem}\label{theorem:A_opt}
Under Restriction \ref{assumption:analogcombiner}, the optimal analog combining matrix, denoted $\mathbf{A}^{\circ}$, is
\begin{align}
    \mathbf{A}^{\circ} = \mathbf{V}_{\mathrm{opt}}^T\boldsymbol{\Sigma}_{\mathbf{x}}^{-\frac{1}{2}},
\end{align}
where the rows of $\mathbf{V}_{\mathrm{opt}}^T\in \mathbb{R}^{P\times N}$ are the $P$ right singular vectors of $\boldsymbol{\tilde{\Gamma}} = \boldsymbol{\Gamma}\boldsymbol{\Sigma}_{\mathbf{x}}^{\frac{1}{2}}$ corresponding to the $P$ largest singular values. The optimal digital processing matrix for a given $\mathbf{A} = \mathbf{A}^\circ$, denoted $\mathbf{D}^\circ(\mathbf{A}^\circ)$, is
\begin{align}
  \mathbf{D}^\circ(\mathbf{A}^\circ) =  \boldsymbol{\Gamma}\boldsymbol{\Sigma}_{\mathbf{x}}^{\frac{1}{2}}\mathbf{V}_{\mathrm{opt}}.
\end{align}
Using $\mathbf{A}^\circ$ and $\mathbf{D}^\circ$ gives the following quantizer-dependent MSE:
\begin{align}\label{eq:quant_dependent_MSE}
    &\mathbb{E}\{||\mathbf{\tilde{s}} - \mathbf{\hat{s}}||^2\} \nonumber\\
    &\quad= \begin{cases}\sum_{i = 1}^{K}\lambda_{\boldsymbol{\tilde{\Gamma}},i}\cdot\rho_{\mathrm{q}}^{(i)}\;\;\quad\qquad\qquad\quad\;\;,\;\mathrm{if }\;P \geq K\\
    \sum_{i = 1}^{P}\lambda_{\boldsymbol{\tilde{\Gamma}},i}\cdot\rho_{\mathrm{q}}^{(i)} + \sum_{i = P+1}^{K}\lambda_{\boldsymbol{\tilde{\Gamma}},i}\;\;,\;\mathrm{otherwise}
    \end{cases}
\end{align}
where $\lambda_{\boldsymbol{\tilde{\Gamma}},i}$ is the $i$-th eigenvalue of $\boldsymbol{\tilde{\Gamma}}\boldsymbol{\tilde{\Gamma}}^T$ (arranged in descending order).
\end{theorem}
\begin{proof}
See Appendix \ref{proof:A_opt}.
\end{proof}

When the number of quantization levels per quantizer is sufficiently large and $P = K$, the distortion vector $\boldsymbol{\eta}$ becomes negligible, and the estimate of the task vector can be expressed as
\begin{align*}
    \mathbf{\hat{s}} \approx& \mathbf{D}^\circ\mathbf{A}^\circ\mathbf{x}\\
    \approx& \boldsymbol{\Gamma}\boldsymbol{\Sigma}_{\mathbf{x}}^{\frac{1}{2}}\mathbf{V}_{\mathrm{opt}}\mathbf{V}_{\mathrm{opt}}^T\boldsymbol{\Sigma}_{\mathbf{x}}^{-\frac{1}{2}}\mathbf{x}\\
    \approx& \boldsymbol{\Gamma}\mathbf{x} \approx \mathbf{\tilde{s}}
\end{align*}
(i.e. our estimate of the task vector approaches the MMSE estimate). Consequently, the quantizer-dependent MSE term approaches zero since $\rho_{\mathrm{q}}^{(p)}\rightarrow 0$ as $\tilde{M}\rightarrow\infty$. \textcolor{black}{In fact, when $\mathbf{x}$ is specialized to a Gaussian vector, we can use the derived quantizer-dependent MSE expression and the approximation for the distortion factor of Gaussian input to show that \[\mathbb{E}[\|\mathbf{\tilde{s}}-\mathbf{\hat{s}}\|^2] \sim \mathcal{O}\left(1/\tilde{M^c}\right),\] where $c = 2$ for non-uniform quantizer and $c = 1.74$ for uniform quantizer.}

The design of the task-based quantizer in Theorem \ref{theorem:A_opt} has a nice intuition. The optimal analog combiner first applies a whitening filter to the measurement vector $\mathbf{x}$. Then, the matrix $\mathbf{V}_{\mathrm{opt}}^T$ maps the ``whitened'' signal from the \emph{measurement space} to a space with lower number of dimensions. We shall call this the \emph{task space}. Quantization is performed in the task space to get $\mathbf{z}$. In the digital domain, the $\boldsymbol{\Sigma}_{\mathbf{x}}^{\frac{1}{2}}\mathbf{V}_{\mathrm{opt}}$ term in $\mathbf{D}^\circ$ inverts the operation of $\mathbf{A}^\circ$ to get some intermediate result $\mathbf{\hat{x}}$, a linear estimate of $\mathbf{x}$ given $\mathbf{z}$. Finally, we compute $\mathbf{\hat{s}} = \boldsymbol{\Gamma}\mathbf{\hat{x}}$ to get an estimate of the task vector.

\begin{figure}[t]
    \includegraphics[scale = .66]{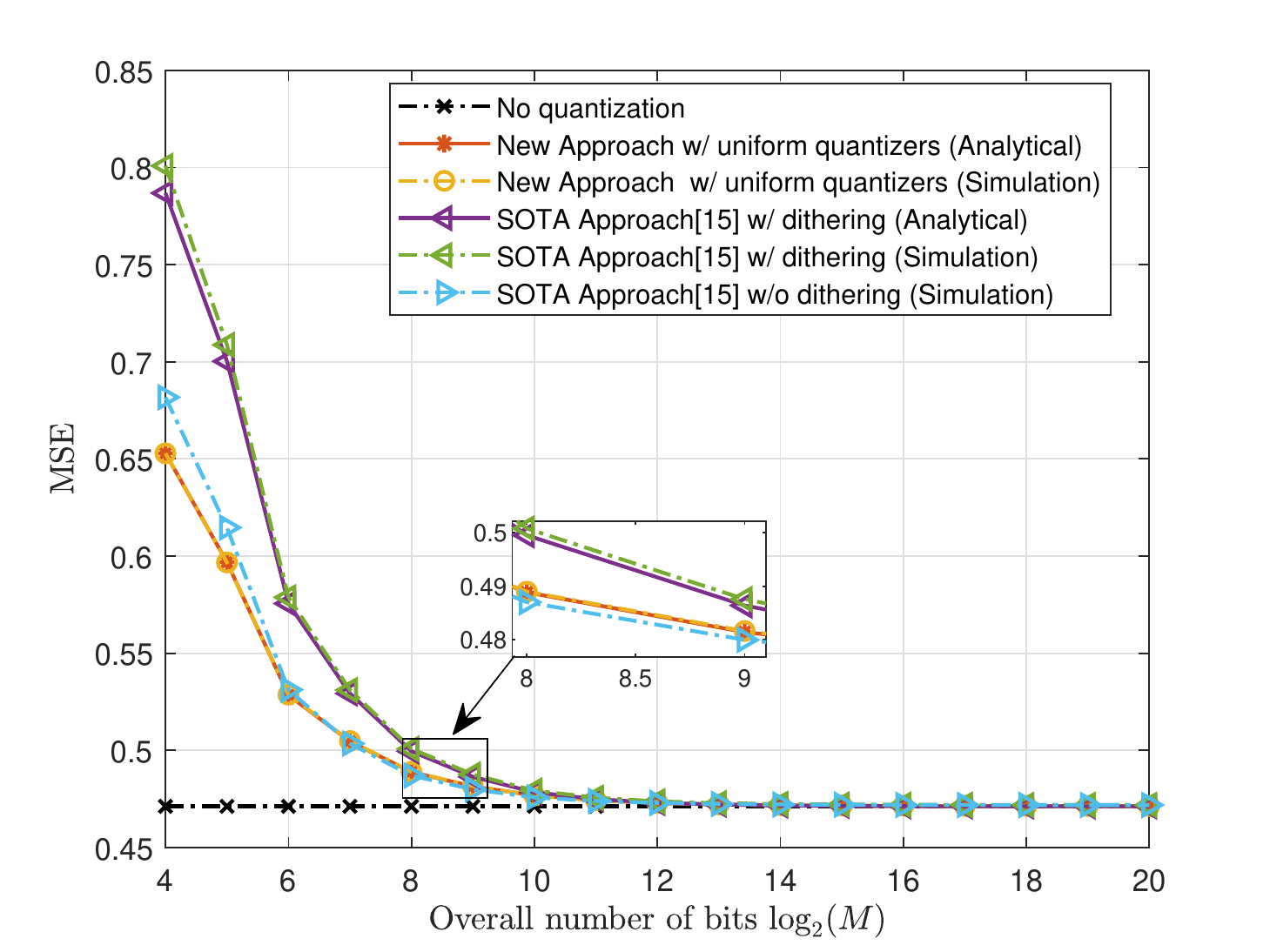}
    \captionsetup{justification=centering}
    \caption{MSE vs overall number of bits of Systems A-D, task: channel estimation with $K = 2$ taps \textcolor{black}{($\sigma_w^2 = 1$)}. }
    \label{fig:channel_estimation_k2}
\end{figure}

\begin{figure}[t]
    \includegraphics[scale = .66]{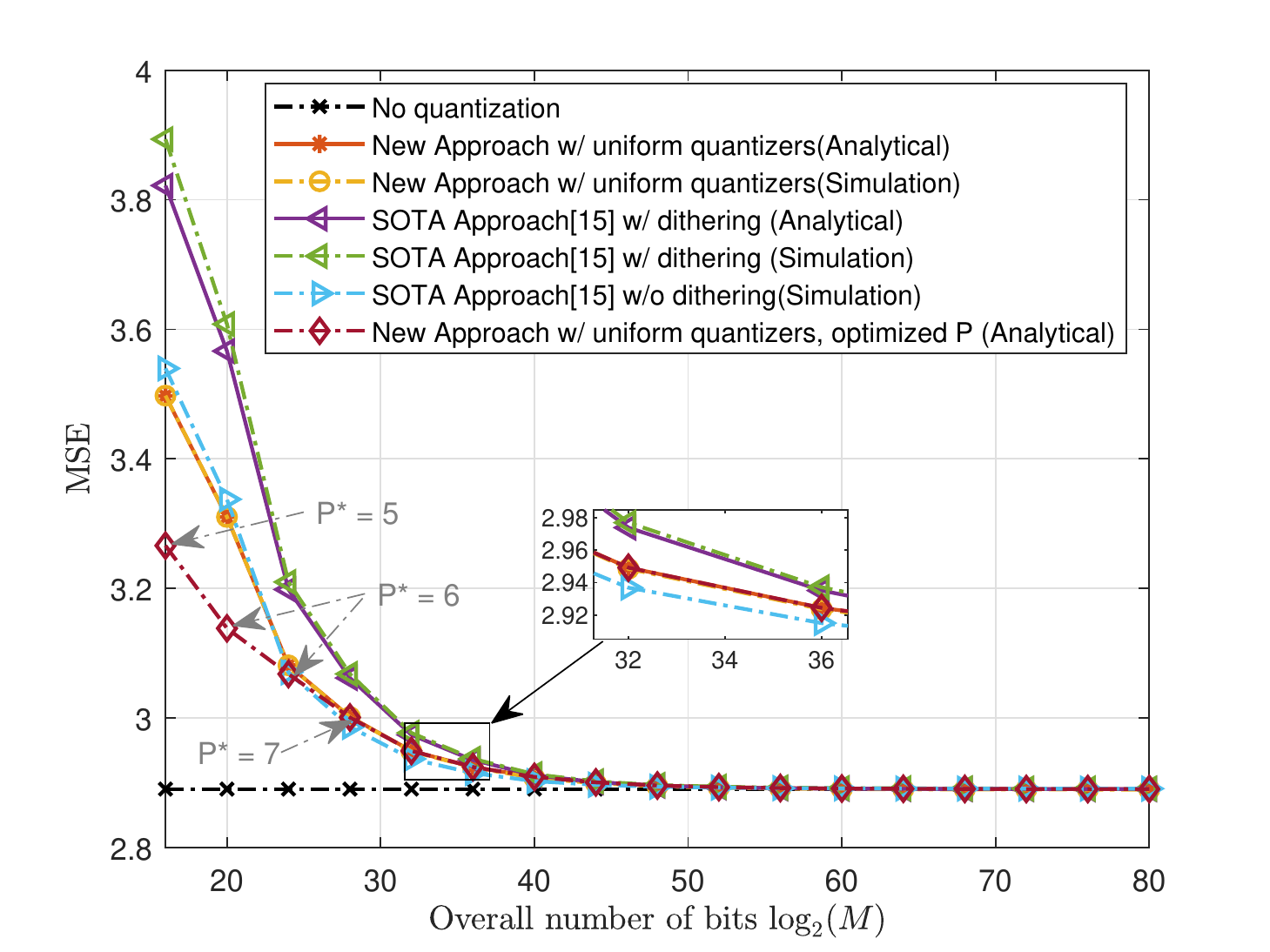}
    \captionsetup{justification=centering}
    \caption{MSE vs overall number of bits of Systems A-D, task: channel estimation with $K = 8$ taps \textcolor{black}{($\sigma_w^2 = 1$)}.}
    \label{fig:channel_estimation_k8}
\end{figure}

There are several differences between the linear mappings and MSE expressions derived in Theorem \ref{theorem:A_opt}, and their counterparts in \cite[Theorem 1]{Shlezinger:2019}. First, the linear mappings we derived are independent of the actual structure of the scalar quantizers and their parameters. The linear mappings only depend on the underlying system task $\boldsymbol{\Gamma}$ and the statistics of the measurement vector $\mathbf{x}$. Thus, we get the same analog linear mappings for both uniform and non-uniform quantizers. In contrast, the optimal linear mappings in \cite[Theorem 1]{Shlezinger:2019} change as the quantizer parameters (e.g. $M$, dynamic range, spacing, etc) are varied. Second, we looked for the optimal $\mathbf{A}$ within the class of analog combiners that satisfy Restriction \ref{assumption:analogcombiner}. However, we have not shown that there is no loss of optimality if we restrict the search space within this class. In fact, the analog linear mappings obtained using the SOTA approach do not necessarily satisfy this property. Third, the quantizer-dependent MSE expressions have different structures. The quantizer-dependent MSE expression in our new approach is a linear combination of the eigenvalues of $\boldsymbol{\tilde{\Gamma}}\boldsymbol{\tilde{\Gamma}}^T$, weighed by the distortion factors of the scalar quantizers. On the other hand, the quantizer-dependent MSE in \cite[Theorem 1]{Shlezinger:2019} and the eigenvalues of $\boldsymbol{\tilde{\Gamma}}\boldsymbol{\tilde{\Gamma}}^T$ exhibit a nonlinear relationship. We provide a more in-depth comparison of the two analysis frameworks in the next section.

\section{Numerical Study for Linear Task}\label{section-numerical_linear}

We now apply our proposed analysis framework for the hardware-limited task-based quantization in a scenario which involves parameter acquisition from quantized observations. More precisely, we consider a scalar channel estimation problem where samples are corrupted by intersymbol interference (ISI) and noise, as in \cite[Section VI-A]{Shlezinger:2019}. The task vector $\mathbf{s}$ represents the coefficients of a $K$-taps multipath channel that we want to estimate. We aim to recover the task vector $\mathbf{s}$ from the $N = 120$ noisy observations contained in $\mathbf{x}$, where the $n$-th element of $\mathbf{x}$ is given by
\begin{align}
    x_n = \sum_{l = 1}^{K}\mathbf{s}_{l} a_{n - l + 1} + w_{n}\quad\forall n\in\{1,2,\cdots,N\}.
\end{align}
The coefficients $\{a_{l}\}$ account for a deterministic training sequence that is known by the task-based quantizer. The quantities $\{w_n\}_{n = 1}^{N}$ represent the i.i.d. zero-mean Gaussian noise process that has unit variance\textcolor{black}{, i.e. $\sigma^2_{w} = 1$,} and is independent of $\mathbf{s}$. The channel $\mathbf{s}$ is modeled as a zero-mean Gaussian vector with the $i$-th row and $j$-th column of its covariance matrix is given by
\begin{align*}
    \Sigma^{(i,j)}_{\mathbf{s}} = e^{-|i-j|},\qquad \forall i,j\in\{1,2,\cdots,K\}.
\end{align*}
Effectively, $\mathbf{x}$ and $\mathbf{s}$ are jointly Gaussian so the linear task assumption $\mathbf{\tilde{s}} = \boldsymbol{\Gamma}\mathbf{x}$ is satisfied, where $\boldsymbol{\Gamma} = \boldsymbol{\Sigma_{\mathbf{s}\mathbf{x}}}\boldsymbol{\Sigma}_{\mathbf{x}}^{-1}$. Finally, we set the training sequence to be
\begin{align}
    a_{l} = \begin{cases}
    \cos\left(\frac{2\pi l}{N}\right),\qquad l > 0\\
    \quad\;\;0\;\;\;\quad,\qquad\mathrm{otherwise}
    \end{cases}.
\end{align}

Using the above setup, we evaluate the distortion of the hardware-limited task-based quantizer designed using our proposed analysis framework, and compare it to that of the hardware-limited task-based quantizer designed using the SOTA analysis framework. We consider two channels: (a) one with $K = 2$ channel taps, and (b) one with $K = 8$ channel taps. By default, we set $P = K$. However, we allow $P$ to be optimized in some parts of the numerical study. For our proposed framework, we used the distortion factors for a Lloyd-Max non-uniform quantizer with Gaussian input. Since $\mathbf{x}$ is a Gaussian random vector, the quantizer input $\mathbf{A}_{p}\mathbf{x}$ for some $p \in \{1,2,\cdots,P\}$ is a linear combination of $N$ Gaussian random variables. Thus, $\mathbf{A}_{p}\mathbf{x}$ is also Gaussian. We also set the range of the overall quantization levels to be $\log_2 M \in [2\cdot K, 10\cdot K]$. Our numerical study will evaluate the distortions incurred by the following quantization systems:

\begin{figure}[t]
    \includegraphics[scale = .66]{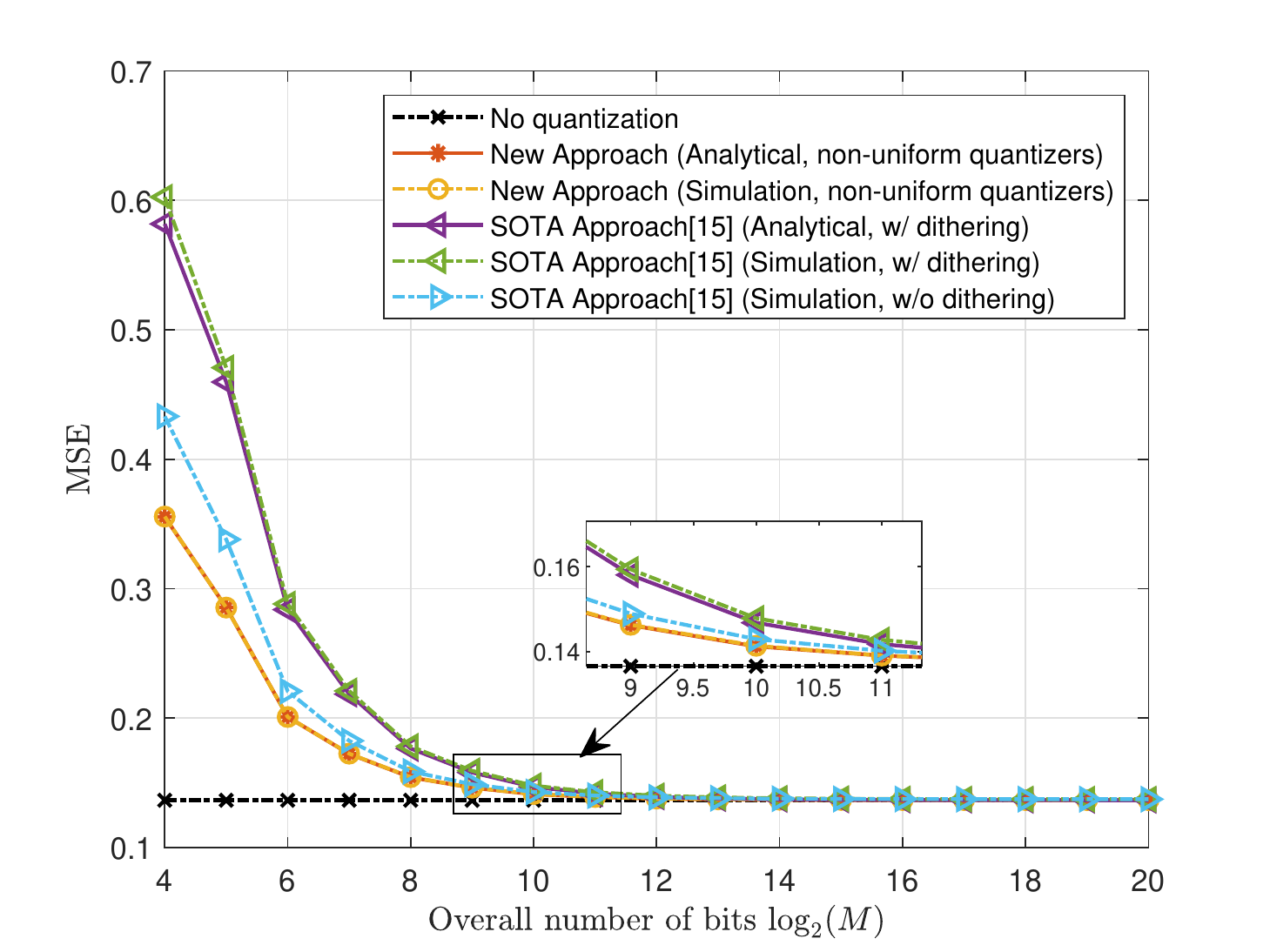}
    \captionsetup{justification=centering}
    \caption{\textcolor{black}{MSE vs overall number of bits of Systems A-D, task: channel estimation with $K = 2$ taps ($\sigma_w^2 = 0.1$)}. }
    \label{fig:channel_estimation_k2_10dB}
\end{figure}

\begin{figure}[t]
    \includegraphics[scale = .66]{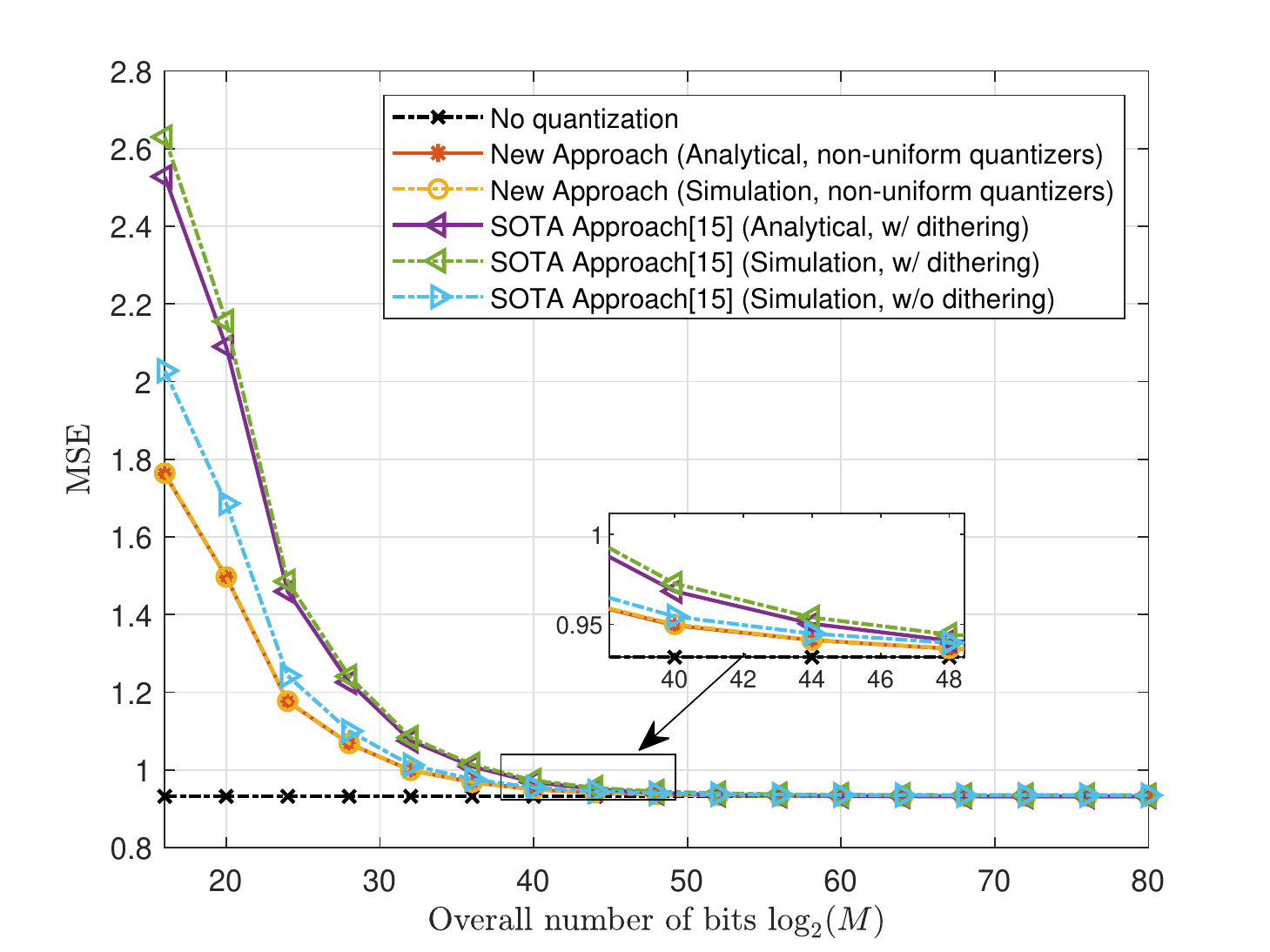}
    \captionsetup{justification=centering}
    \caption{\textcolor{black}{MSE vs overall number of bits of Systems A-D, task: channel estimation with $K = 8$ taps ($\sigma_w^2 = 0.1$)}.}
    \label{fig:channel_estimation_k8_10dB}
\end{figure}

\begin{figure}[t]
    \includegraphics[scale = .66]{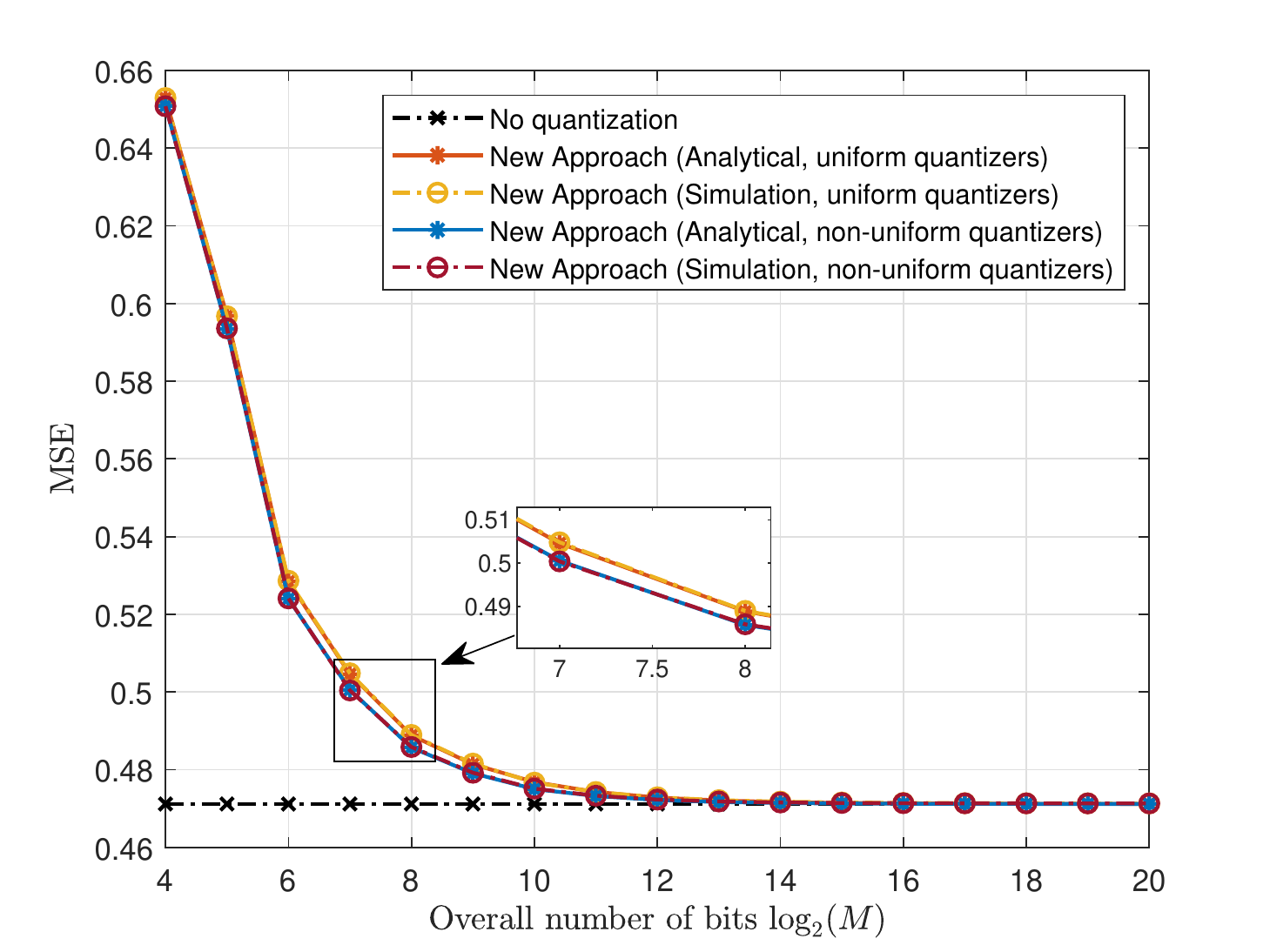}
    \captionsetup{justification=centering}
    \caption{Comparison of task-based system with uniform and non-uniform quantization \textcolor{black}{($K = 2, \sigma_w^2 = 1$)}.}
    \label{fig:uniform_vs_nonuniform}
\end{figure}

\begin{figure*}[t]
    \centering
    \includegraphics[scale = .725]{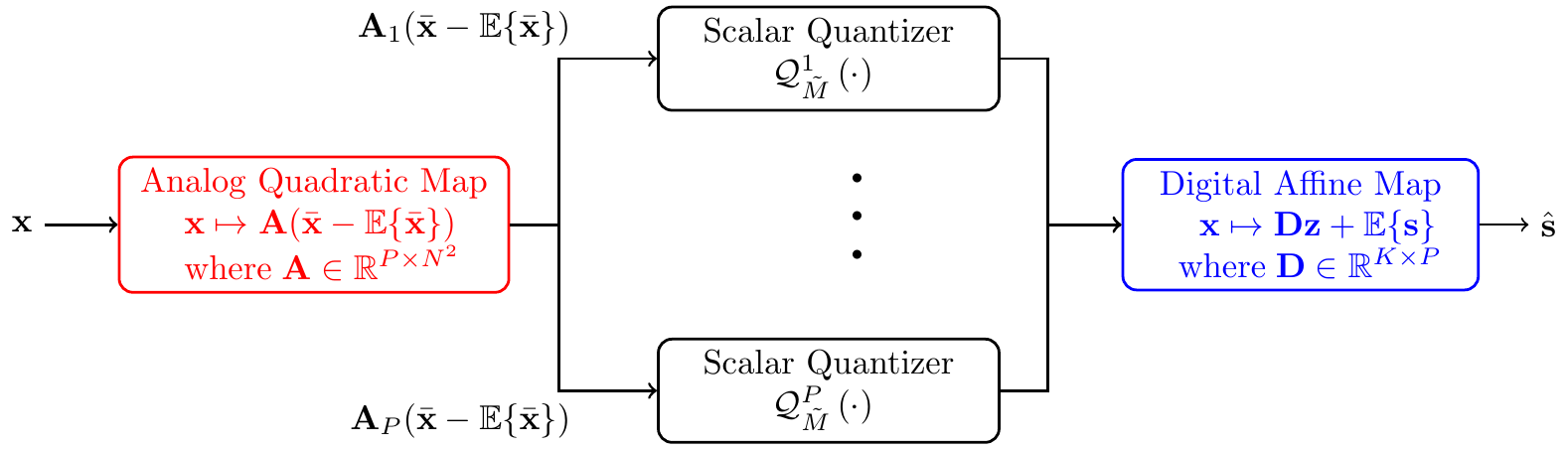}
    \captionsetup{justification=centering}
    \caption{System model of task-based quantization with analog quadratic mapping and digital affine mapping.}
    \label{fig:tb_quadratic}
\end{figure*}

\begin{itemize}
    \item \textbf{System A (No quantization):} The MMSE $\mathbb{E}\{||\mathbf{s} - \tilde{\mathbf{s}}||^2\}$, which is the optimal distortion of an unquantized system. This quantity can be expressed as
    \begin{align}\label{eq:unquantized}
        \mathbb{E}\{||\mathbf{s} - \tilde{\mathbf{s}}||^2\} = \mathrm{Tr}\left(\boldsymbol{\Sigma}_{\mathbf{s}}\right) - \mathrm{Tr}\left(\boldsymbol{\Gamma}\boldsymbol{\Sigma}_{\mathbf{s}\mathbf{x}}^T\right).
    \end{align}
    \item \textbf{System B (SOTA Approach w/ dithering):} This is the distortion of the dithered hardware-limited task-based quantizer in which the analog and digital linear mappings $\mathbf{A}$ and $\mathbf{D}$ are designed using \cite[Theorem 1]{Shlezinger:2019}. Both simulated and theoretical distortions are evaluated. The simulated MSEs are computed empirically by averaging the MSE over 500,000 Monte Carlo runs. 
    \item \textbf{System C (SOTA Approach w/o dithering):} Since dithering increases the energy of the quantization noise, we also simulate the MSE of the hardware-limited task-based quantizer without dithering ($\mathbf{A}$ and $\mathbf{D}$ are still designed using \cite[Theorem 1]{Shlezinger:2019}).
    \item \textbf{System D (New Approach, uniform quantizers):} This is the distortion of the hardware-limited task-based quantizer designed under our proposed analysis framework. The uniform quantizers are designed using the Lloyd-Max algorithm for equally-spaced level quantizers (See \cite[Equation 8]{Max:1960}). Both simulated and theoretical distortions are evaluated. The simulated MSEs are computed empirically by averaging the MSE over 500,000 Monte Carlo runs. Moreover, our analytical expression enables us to optimize the number of scalar quantizers. We also present the theoretical MSE using the optimal $P$, denoted $P^*$. This is computed by trying all possible $P \in \{1,2,\cdots, K\}$ in \eqref{eq:quant_dependent_MSE}.
    \item \textbf{System E (New Approach, non-uniform quantizers)}: This is the same as System D but we allow the quantizers to be non-uniform. The thresholds and representative levels of the scalar quantizers are designed using the Lloyd-Max algorithm \cite{Max:1960}. Note that we did not change the configuration of the linear mappings since the derived linear mappings in Theorem \ref{theorem:A_opt} are agnostic of the actual structure of the scalar quantizers. %\textcolor{black}{Note that $\mathbf{A}$ and $\mathbf{D}$ in System D and System E are the same since they do not depend on the exact structure of the scalar quantizers.}
\end{itemize}

Figures \ref{fig:channel_estimation_k2} and \ref{fig:channel_estimation_k8} depict the distortions for System A to System D for $K = 2$ and $K = 8$ channel taps, respectively. In both cases, it can be observed that the MSE of System D is lower than that of System B. The performance gain is more pronounced in the low resolution regime but the gap between the MSEs of the two frameworks diminishes as the overall number of bits is increased. When all the scalar quantizers in the quantizer model have at least five bits, i.e. $\log_2 M \geq 5K$, the quantizer-dependent MSE is negligible and most of the overall MSE comes from \eqref{eq:unquantized}. We also demonstrate in the $K = 8$ setup that using lower $P$ may yield lower MSE when there is a tight quantization budget. 

There is no clear winner between System C and System D. System D has lower MSE when the overall number of bits are limited but is slightly outperformed by System C at some values of $\log_2(M)$. We conjecture that its subpar performance at some cases is due to Restriction \ref{assumption:analogcombiner}. That is, there is loss of optimality when restricting the search for the optimal $\mathbf{A}$ within a class of analog linear mappings that satisfy Restriction \ref{assumption:analogcombiner}. \textcolor{black}{In fact, we inspected the analog linear maps in System C and noticed that $\mathbf{A}\boldsymbol{\Sigma}_{\mathbf{x}}\mathbf{A}^T$ is not a diagonal matrix, thus violating Restriction 1.} Nonetheless, we point out that the simulated MSE and the theoretical MSE (i.e. Equation \eqref{eq:unquantized} + Equation \eqref{eq:quant_dependent_MSE}) of System D perfectly coincide in our numerical study. This is expected since the proposed framework is exact, provided the assumptions on the scalar quantizers and analog combining matrix are satisfied. On the other hand, we can see that the simulated MSE of the dithered task-based quantizer designed using the SOTA framework does not perfectly match the theoretical MSE. This is because the overload probabilities of the quantizers are nonzero. Thus, \cite[Theorem 1]{Shlezinger:2019} only holds approximately. Furthermore, the SOTA analysis framework is not capable of accurately predicting the simulated MSE of System C.

\textcolor{black}{We also extend the numerical analysis of the channel estimation task in the high signal-to-noise ratio regime. More specifically, we compare the performances of the System C and System D when the noise variance in the numerical study is set to $\sigma_{w}^2 = 0.1$. The numerical results for $K = 2$ and $K = 8$ channel taps are depicted in Figures \ref{fig:channel_estimation_k2_10dB} and \ref{fig:channel_estimation_k8_10dB}, respectively. It can be seen that the MSE of System D is now consistently lower than that of System C for all quantization bit budgets $\log_2(M)$ considered in the numerical study. Furthermore, our theoretical prediction for the MSE of System D still coincides with the simulated MSE. Overall, these findings suggest that the task-based quantizer should be designed using our proposed framework rather than the SOTA framework when the energy of the additive noise embedded in the observations is small or when the quantization bit budget is limited.}

When the scalar quantizers in System D are replaced with non-uniform quantizers, i.e. System E, we observe in Figure \ref{fig:uniform_vs_nonuniform} that the MSE of the task-based quantizer designed using our proposed analysis framework slightly improved. The use of non-uniform quantizers in our proposed framework is expected to provide performance gain, albeit small, since non-uniform quantizers generally have lower distortion factor compared to uniform quantizers. More importantly, we emphasize that the simulated MSE of the task-based system equipped with non-uniform quantizers coincides with our theoretical predictions. The proposed framework enables a model-based analysis of task-based quantization with non-uniform quantizers. To the best of our knowledge, only a data-driven approach \cite{Shlezinger:2021} for task-based quantization with non-uniform quantizers is available in the literature.

In the next section, we extend the framework to quadratic tasks.

\section{Extension of the Proposed Framework to Quadratic Task}\label{section-quadtask}

\begin{figure*}[t]
    %\centering
    \hspace*{-2.25cm}
    \includegraphics[scale = .725]{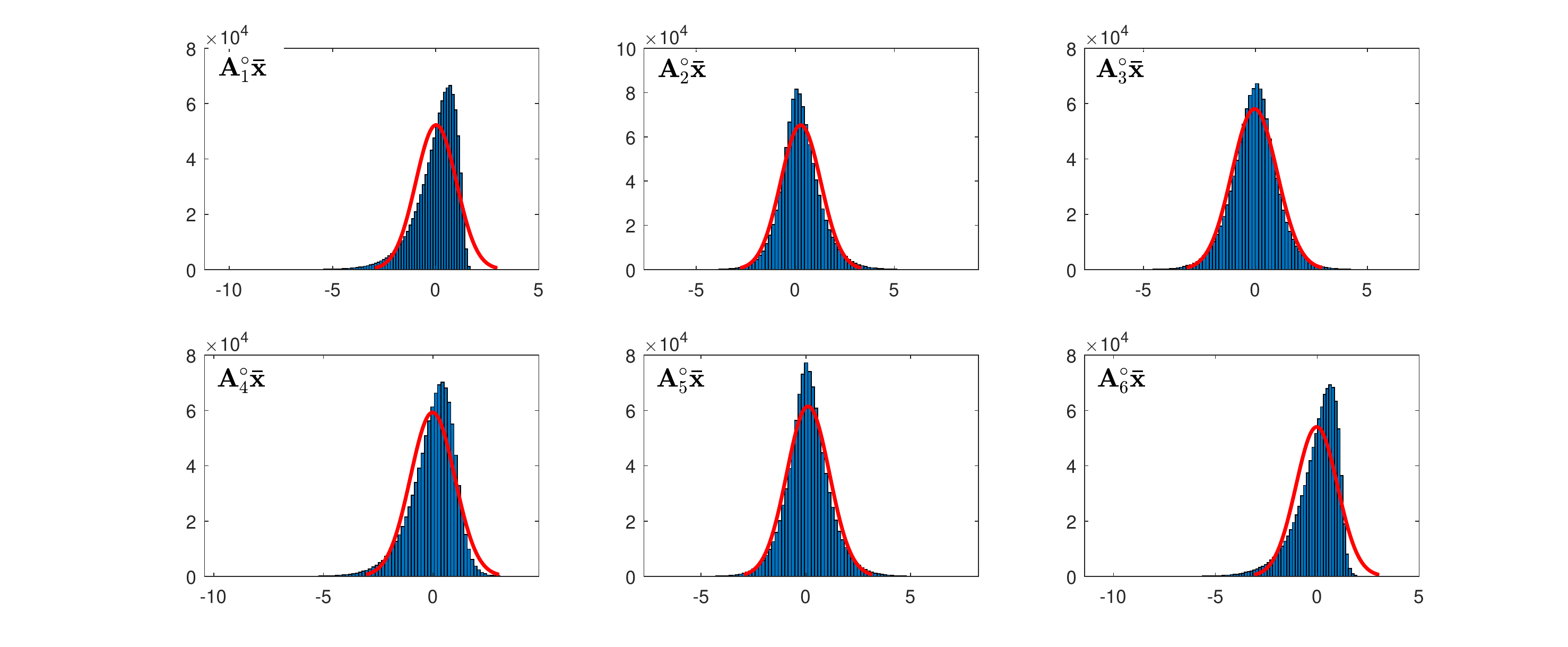}
    \captionsetup{justification=centering}
    \caption{Empirical distributions of the six elements of $\mathbf{A}^\circ\bar{\mathbf{x}}$. The distributions are generated from 1,000,000 Monte Carlo runs. The red plots correspond to the best-fit Gaussian distributions.}
    \label{fig:distribution_analog_out}
\end{figure*}

\subsection{Setup for the Quadratic Task Problem}

To extend the proposed analysis framework to the quadratic task problem, we consider the task-based quantization model depicted in Figure \ref{fig:tb_quadratic} and follow the approach of \cite{Salamatian:2019}. The measurement vector $\mathbf{x} \in \mathbb{R}^{N\times 1}$ is a zero-mean Gaussian random vector and the task is to recover a set of quadratic functions $\{\mathbf{x}^T\mathbf{C}_{k}\mathbf{x}\}_{k = 1}^{K}$, where each $\mathbf{C}_{k} \in\mathbb{R}^{N\times N}$ satisfies $\mathbb{E}\left\{\mathbf{x}^T\mathbf{C}_{k}\mathbf{x}\right\} < \infty$. We shall represent the results of these quadratic functions using a $K\times 1$ task vector $\mathbf{s}$ whose entries are given by $s_{k} = \mathbf{x}^T\mathbf{C}_{k}\mathbf{x}$. 

Since we are interested in quadratic tasks, we introduce the quadratic measurement vector $\mathbf{\bar{x}}= \mathrm{vec}\left(\mathbf{x}\mathbf{x}^T\right) \in\mathbb{R}^{N^2\times 1}$, where $\mathrm{vec}(\mathbf{C}) \in \mathbb{R}^{mn\times 1}$ is the vectorization of $\mathbf{C}^{m\times n}$, i.e. the vector is a vertical stacking of the columns of $\mathbf{C}$. We also introduce $\mathbf{G} \in \mathbb{R}^{K\times N^2}$ whose $k$-th row is given by $\mathrm{vec}^T(\mathbf{C}_{k})$. Consequently, we can write the task vector as $\mathbf{s} = \mathbf{G}\mathbf{\bar{x}}$. \textcolor{black}{The following proposition from \cite{Salamatian:2019} gives the structure of the analog and digital mappings for the quadratic task.
\begin{proposition}(\cite[Theorem 2]{Salamatian:2019})\label{proposition:quad_struct}
For any $P\times N^2$ matrix $\mathbf{A}$ with $P \leq N^2$, the MMSE estimate of a quadratic function $f(\mathbf{x}) = \mathbf{x}^T\mathbf{C}\mathbf{x}$ from random vector $\mathbf{W} = \mathbf{A}(\mathbf{\bar{x}} - \mathbb{E}[\mathbf{\bar{x}}])$ can be written as
    \begin{align*}
        \mathbf{E}[f(\mathbf{x})|\mathbf{W}] = \mathbf{d}^T\mathbf{W} + \mathbb{E}[f(\mathbf{x})]
    \end{align*}
    for some $P\times 1$ vector $\mathbf{d}$, which depends on $\mathbf{C}$, $\mathbf{A}$, and the covariance of $\mathbf{x}$.
\end{proposition}
Due to Proposition \ref{proposition:quad_struct}}, we can focus on analog mapping $h_{\mathrm{a}}$ of the form 
\begin{align}
h_{\mathrm{a}}:\quad \mathbf{x} \mapsto \mathbf{A}\left(\mathbf{\bar{x}} - \mathbb{E}\{\mathbf{\bar{x}}\}\right),    
\end{align}
where $\mathbf{A} \in \mathbb{R}^{P\times N^2}$ is a matrix that applies a rotation and dimensionality reduction to the shifted quadratic measurement vector $\mathbf{\bar{x}} - \mathbb{E}\{\mathbf{\bar{x}}\}$, and on digital mapping $h_{\mathrm{d}}$ of the form
\begin{align}
  h_{\mathrm{d}}:\quad \mathbf{x} \mapsto \mathbf{D}\mathbf{z} + \mathbb{E}\{\mathbf{s}\},      
\end{align}
where $\mathbf{z}\in \mathbb{R}^{P\times 1}$ contains the $P$ outputs of the scalar quantizers. We find the matrices $\mathbf{A}$ and $\mathbf{D}$ that minimizes $\mathbb{E}\{||\mathbf{s} - \mathbf{\hat{s}}||^2\}$, where $\mathbf{\hat{s}}$ is the output of the task-based quantizer.

%This formulation ``linearizes'' the quadratic task problem. \textcolor{black}{The following proposition }

\subsection{System Design}
We now apply our proposed framework to the \emph{linearized} quadratic task. First, we let $\boldsymbol{\Sigma}_{\mathbf{\bar{x}}} \in \mathbb{R}^{N^2\times N^2}$ be the covariance matrix of $\mathbf{\bar{x}}$.  Since $\mathbf{x}$ is a zero-mean Gaussian random vector, $\mathbf{x}\mathbf{x}^T$ is an $N\times N$ Wishart matrix of degree 1. Thus, the elements of $\boldsymbol{\Sigma}_{\mathbf{\bar{x}}} \in \mathbb{R}^{N^2\times N^2}$ and $\mathbf{E}\{\mathbf{\bar{x}}\}$ can be obtained from \cite{Nydick:2012}. Alternatively, these quantities can be computed empirically as done in \cite{Salamatian:2019}. The following corollary of Theorem \ref{theorem:A_opt} gives the quadratic task extension of the proposed framework.

\begin{corollary}\label{corollary:quad_task}
Under Restriction \ref{assumption:analogcombiner}, the optimal analog combining matrix, denoted $\mathbf{A}^{\circ}$, is
\begin{align}
    \mathbf{A}^{\circ} = \bar{\mathbf{V}}_{\mathrm{opt}}^T\boldsymbol{\Sigma}_{\mathbf{\bar{x}}}^{-\frac{1}{2}},
\end{align}
where the rows of $\bar{\mathbf{V}}_{\mathrm{opt}}^T\in \mathbb{R}^{P\times N^2}$ are the $P$ right singular vectors of $\mathbf{\tilde{G}} = \mathbf{G}\boldsymbol{\Sigma}_{\mathbf{\bar{x}}}^{\frac{1}{2}}$ corresponding to the $P$ largest singular values. The optimal digital processing matrix for a given $\mathbf{A} = \mathbf{A}^\circ$, denoted $\mathbf{D}^\circ(\mathbf{A}^\circ)$, is
\begin{align}
  \mathbf{D}^\circ(\mathbf{A}^\circ) =  \mathbf{G}\boldsymbol{\Sigma}_{\mathbf{\bar{x}}}^{\frac{1}{2}}\bar{\mathbf{V}}_{\mathrm{opt}}.
\end{align}
Using $\mathbf{A}^\circ$ and $\mathbf{D}^\circ$ gives the following MSE:
\begin{align}\label{eq:quant_dependent_MSE_quadtask}
    &\mathbb{E}\{||\mathbf{s} - \mathbf{\hat{s}}||^2\} \nonumber\\
    &\quad= \begin{cases}\sum_{i = 1}^{K}\lambda_{\mathbf{\tilde{G}},i}\cdot\rho_{\mathrm{q}}^{(i)}\;\;\quad\qquad\qquad\quad\;\;,\;\mathrm{if }\;P \geq K\\
    \sum_{i = 1}^{P}\lambda_{\mathbf{\tilde{G}},i}\cdot\rho_{\mathrm{q}}^{(i)} + \sum_{i = P+1}^{K}\lambda_{\mathbf{\tilde{G}},i}\;,\;\mathrm{otherwise}
    \end{cases}
\end{align}
where $\lambda_{\mathbf{\tilde{G}},i}$ is the $i$-th eigenvalue of $\mathbf{\tilde{G}}\mathbf{\tilde{G}}^T$ (arranged in descending order).
\end{corollary}
\begin{proof}
\textcolor{black}{
The corollary directly follows from Theorem \eqref{theorem:A_opt} since $\mathbf{s}$ is a linear function of $\mathbf{\bar{x}}$, i.e. $\mathbf{s} = \mathbf{G}\mathbf{\bar{x}}$.}
\end{proof}

\subsection{Numerical Results}
We now demonstrate the effectiveness of the proposed framework on the quadratic task problem. We consider the empirical covariance estimation problem described in \cite[Section V]{Salamatian:2019}. The input is given by $\mathbf{x} = [\mathbf{y}_1^T,\cdots,\mathbf{y}_4^T]^T$, where $\{\mathbf{y}_l\}_{l = 1}^{4}$ are i.i.d. 3$\times$1 zero mean Gaussian random vectors. Hence, the measurement vector $\mathbf{x}$ is a 12$\times$1 vector. The $i$-th row and $j$-th column of the covariance matrix of each $\mathbf{y}_{l}$, denoted $\boldsymbol{\Sigma}_{\mathbf{y}}$, is given by
\begin{align*}
    \boldsymbol{\Sigma}_{\mathbf{y}}^{(i,j)} = e^{-|i-j|},\qquad\forall i,j \in \{1,2,3\}.
\end{align*}
The parameter we want to recover is a 3$\times$3 empirical covariance matrix $\frac{1}{4}\sum_{l = 1}^{4}\mathbf{y}_l\mathbf{y}_l^T$, which is completely determined by its upper triangular matrix. Thus, the task vector $\mathbf{s}$ has length $K = 6$.

Since $\mathbf{\bar{x}}$ is not a Gaussian random vector, we expect $\mathbf{A}^\circ\mathbf{\bar{x}}$ to be non-Gaussian as well. Indeed, as illustrated in Figure \ref{fig:distribution_analog_out}, the $P$ outputs of the analog quadratic mapping are non-Gaussian. Therefore, we use the $\mathrm{lloyds}(\cdot)$ function of MATLAB to get the $P$ Lloyd-Max scalar quantizers and their corresponding distortion factors.

\begin{figure}[t]
    \includegraphics[scale = .66]{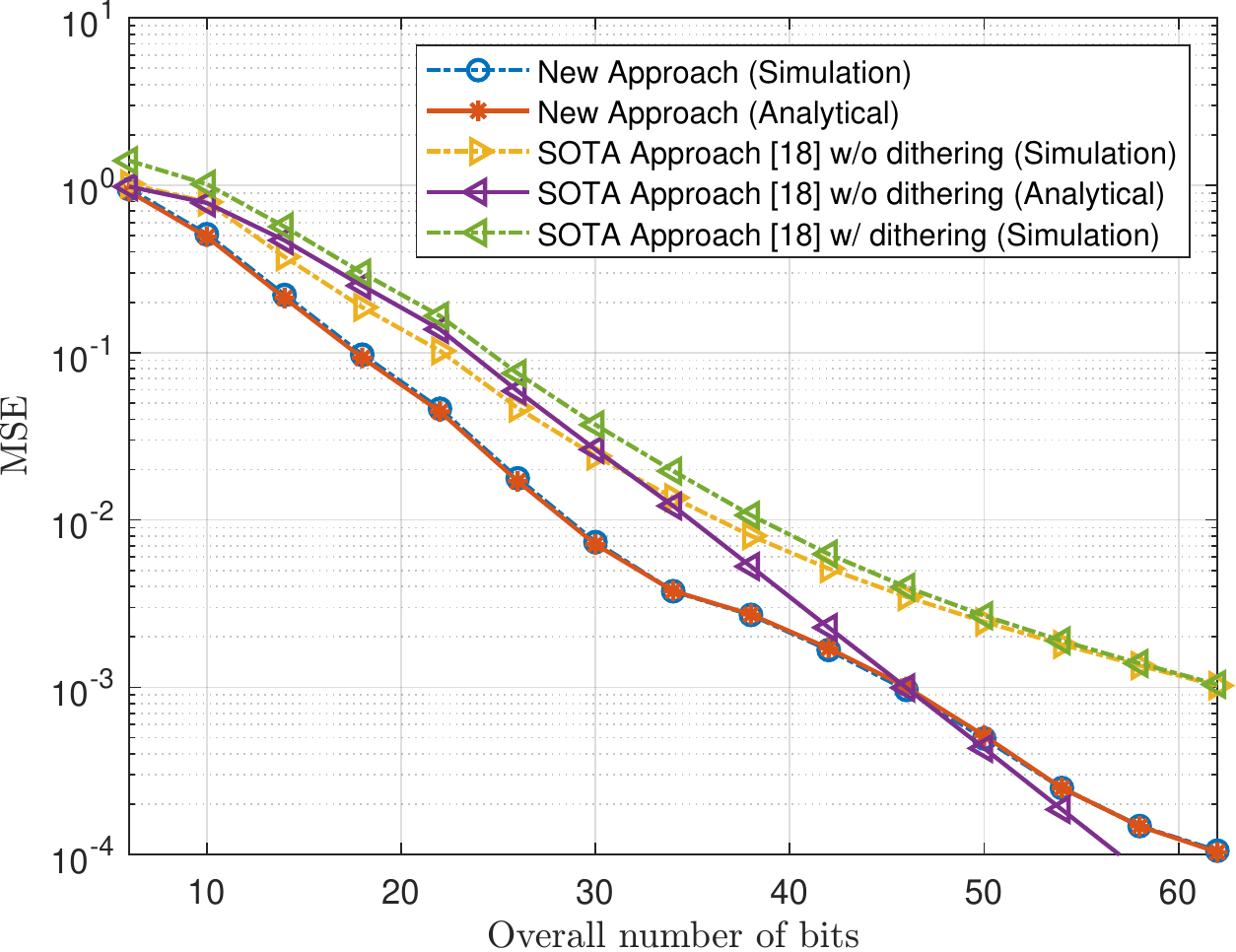}
    \captionsetup{justification=centering}
    \caption{MSE vs overall number of bits, empirical covariance recovery.}
    \label{fig:empirical_cov}
\end{figure}

We evaluate the distortion incurred by the following task-based quantization systems:
%\footnote{In addition to the issues in footnote 1, the proponents of \cite{Salamatian:2019} used an $\eta$ in their code that varies with the number of bits. This is different from what is described in the paper ($\eta = 3$). We followed the former since it closely reproduce their numerical results.}
%\footnote{Since the elements of $\mathbf{A}^\circ\hat{\mathbf{x}}$ are non-Gaussian and non-i.i.d., the distortion factors are first obtained empirically and then applied to equation \eqref{eq:quant_dependent_MSE_quadtask}.}
\begin{itemize}
    \item \textbf{System F (SOTA Approach for Quadratic Task \cite{Salamatian:2019}):} This is the distortion of the task-based quantizer where $\mathbf{A}$ and $\mathbf{D}$ are designed using \cite[Theorem 3]{Salamatian:2019}. We present the simulated MSEs of both the dithered and non-dithered case. The simulated MSEs are computed empirically by averaging the MSE over 500,000 Monte Carlo runs. The theoretical MSE \cite[Theorem 3]{Salamatian:2019} is also evaluated.
    \item \textbf{System G (New Approach for Quadratic Task, non-uniform quantizers):} This is the distortion of the task-based quantizer where $\mathbf{A}$ and $\mathbf{D}$ are designed using Corollary \ref{corollary:quad_task}. Both simulated and analytical MSEs are evaluated. Since the elements of $\mathbf{A}^\circ\bar{\mathbf{x}}$ are non-Gaussian and non-i.i.d., the distortion factors are first obtained empirically and then applied to equation \eqref{eq:quant_dependent_MSE_quadtask} to produce the analytical MSE. The simulated MSE is computed empirically by averaging the MSE over 500,000 Monte Carlo runs.
\end{itemize}

Figure \ref{fig:empirical_cov} shows the distortions of the quantization systems. It can be observed that the simulated MSE of System G is lower than that of non-dithered System F. Moreover, the analytical expression we derived closely matches the simulated MSE of our task-based quantizer. This numerical result demonstrates that our proposed analysis framework can be potentially applied to problems with nonlinear tasks. On the other hand, we see that while the analytical MSE expression in \cite[Theorem 3]{Salamatian:2019} expects System F to yield the lowest MSE in certain scenarios, it does not accurately predict the actual MSE of the dithered task-based quantizer designed using the SOTA framework. In fact, the relative discrepancy gets worse as the number of quantization levels is increased.

% \textcolor{black}{The analytical and simulated MSE of dithered System F are also depicted in Figure \ref{fig:empirical_cov}. While the analytical MSE in \cite[Theorem 3]{Salamatian:2019} displays a really good performance in some scenarios, we note that the simulated MSE of the task-based quantizer designed using \cite[Theorem 3]{Salamatian:2019} is not accurately described by the analytical MSE expression.}

\section{Conclusion}\label{section-conclusion}

In this work, we formulated a new analysis framework based on the Bussgang decomposition for hardware-limited task-based quantization that overcomes limitations of the current SOTA framework. More precisely, our framework does not rely on the zero overload probability assumption and works for both uniform and non-uniform scalar quantizers without dithering. Our first contribution is a rigorous derivation of the optimal linear mappings and analytical MSE in the linear task scenario under a restriction on the analog combiner. In contrast to the linear mappings derived under the SOTA framework, our mappings do not depend on the structure of the scalar quantizers. We then demonstrated in our numerical study that the simulated MSE of the task-based quantizer designed under our proposed framework coincides with our theoretical predictions. Additionally, we also extended our framework to quadratic tasks and showed that our analytical expression for the MSE continues to hold provided that the distortion factors are known or can be computed empirically.

One notable research direction is to investigate the applicability of the analysis framework to nonlinear tasks beyond the quadratic task problem. For instance, can we extend our proposed method to design and analyze hardware-limited task-based quantizers performing classification tasks? It is also interesting to see how the proposed framework can be utilized to design power-efficient analog mappings. 

%\textcolor{black}{Lastly, Restriction \ref{assumption:analogcombiner} is imposed on the analog combiner to obtain analytically tractable results. It is not yet clear when Restriction \ref{assumption:analogcombiner} becomes highly limiting. As such, it is necessary to understand under what conditions does Restriction \ref{assumption:analogcombiner} incur significant performance loss.}

%It is necessary to understand under what conditions does Restriction \ref{assumption:analogcombiner} incur significant performance loss.}

%Although the paper did not dwell on the implementation aspect, our framework can still serve as a new benchmark for evaluating the performance of hardware-limited task-based quantization.

\begin{appendices}
\section{Proof of Proposition \ref{proposition:B_diagonal}}\label{proof:B_diagonal}

By Restriction \ref{assumption:analogcombiner}, $\mathbf{A}\boldsymbol{\Sigma}_{\mathbf{x}}\mathbf{A}^T$ becomes a diagonal matrix. That is,
\begin{align}
  \mathbf{A}\boldsymbol{\Sigma}_{\mathbf{x}}\mathbf{A}^T = \mathrm{diag}\{\mathbf{P}_{\mathbf{A}\mathbf{x}}\},  
\end{align}
where $\mathbf{P}_{\mathbf{A}\mathbf{x}} = [P_{\mathbf{A}\mathbf{x}}^{(1)},\cdots,P_{\mathbf{A}\mathbf{x}}^{(P)}]^T$ and $P_{\mathbf{A}\mathbf{x}}^{(i)} = \mathbb{E}\{(\mathbf{A}_{i}\mathbf{x})^2\}$. In addition, it can be shown that
\begin{align*}
    \mathbb{E}\{z_{i}\mathbf{x}^T\mathbf{A}_{j}^T\} =& \mathbb{E}\{(B_{ii}\mathbf{A}_i\mathbf{x} + \eta_{i})\mathbf{x}^T\mathbf{A}_{j}^T\}\\
    =&B_{ii}\mathbf{A}_{i}\boldsymbol{\Sigma}_{\mathbf{x}}\mathbf{A}_{j}^T + \mathbb{E}\{\eta_{i}\mathbf{x}^T\mathbf{A}_j^T\}\\
    =& 0
\end{align*}
for $i \neq j$. The quantity $B_{ii}$ is the $i$-th diagonal entry of $\mathbf{B}$. The first line follows by applying Bussgang decomposition at the output of the $i$-th quantizer. The first term in the second line is zero due to Restriction \ref{assumption:analogcombiner} while the second term in the second line is zero since the distortion at the $i$-th quantizer is uncorrelated with the input of the $j$-th quantizer. Thus, $\boldsymbol{\Sigma}_{\mathbf{z}\mathbf{x}}\mathbf{A}^T$ is a diagonal matrix. Consequently, the Bussgang gain matrix in \eqref{eq:Bussgang_gain} is also diagonal.

To derive the diagonal elements of the Bussgang gain matrix, we expand \eqref{definition:rho_q}:
\begin{align*}
    \rho_{\mathrm{q}}^{(p)} =& \frac{\mathbb{E}\{(z_{p} - \mathbf{A}_{p}\mathbf{x})^2\}}{\mathbb{E}\{(\mathbf{A}_{p}\mathbf{x})^2\}}\\
    =& \frac{\mathbb{E}\{z_{p}^2\} - 2\mathbb{E}\{z_{p}\mathbf{x}^T\mathbf{A}_{p}^T\} + \mathbb{E}\{(\mathbf{A}_{p}\mathbf{x})^2\}}{\mathbb{E}\{(\mathbf{A}_{p}\mathbf{x})^2\}}.
\end{align*}
Note that we considered scalar quantizers that satisfy the property $\mathbb{E}\{X_{\mathrm{in}}|\mathcal{Q}(X_{\mathrm{in}})\} = \mathcal{Q}(X_{\mathrm{in}})$. As such, we have
\begin{align}\label{eq:covariance_Z}
    \mathbb{E}\{z_{p}\mathbf{x}^T\mathbf{A}_p^T\} =& \mathbb{E}\{\mathbb{E}\{z_{p}\mathbf{x}^T\mathbf{A}_p^T|z_{p}\} \}\nonumber\\
    =& \mathbb{E}\{z_{p}^2\},
\end{align}
where the first line follows from the law of iterated expectation, and the second line follows from $\mathbb{E}\{X_{\mathrm{in}}|\mathcal{Q}(X_{\mathrm{in}})\} = \mathcal{Q}(X_{\mathrm{in}})$. Effectively, the distortion factor becomes
\begin{align*}
    \rho_{\mathrm{q}}^{(p)} =& \frac{ \mathbb{E}\{(\mathbf{A}_{p}\mathbf{x})^2\} - \mathbb{E}\{z_{p}\mathbf{x}^T\mathbf{A}_{p}^T\}}{\mathbb{E}\{(\mathbf{A}_{p}\mathbf{x})^2\}}\\
    =& 1 - \frac{\mathbb{E}\{z_{p}\mathbf{x}^T\mathbf{A}_{p}^T\}}{\mathbb{E}\{(\mathbf{A}_{p}\mathbf{x})^2\}}\\
    =& 1 - B_{pp}.
\end{align*}
The claim is proven by doing the above analysis for all $p \in \{1,\cdots, P\}$.

\section{Proof of Proposition \ref{proposition:D_opt}}\label{proof:D_opt}

From \eqref{eq:MSE_objectivefunc}, we can simply focus on finding $\mathbf{D}^\circ$ that minimizes the quantizer-dependent MSE for a given $\mathbf{A}$. Under the assumption that $\tilde{\mathbf{s}}$ is a linear task, the optimal $\mathbf{D}$ which results in $\hat{\mathbf{s}}$ being the linear MMSE estimate of $\tilde{\mathbf{s}}$ given $\mathbf{z} = \mathbf{B}\mathbf{A}\mathbf{x} + \boldsymbol{\eta}$ is
\begin{align}\label{eq:B_opt_derivation}
    \mathbf{D}^{\circ}(\mathbf{A}) =& \mathbb{E}\{\tilde{\mathbf{s}}\mathbf{z}^T\}\mathbb{E}\{\mathbf{z}\mathbf{z}^T\}^{-1}\nonumber\\
    =& \mathbb{E}\{\tilde{\mathbf{s}}(\mathbf{B}\mathbf{A}\mathbf{x} + \boldsymbol{\eta})^T\}\mathbb{E}\{\mathbf{z}\mathbf{z}^T\}^{-1}\nonumber\\
    =& \left(\boldsymbol{\Gamma}\boldsymbol{\Sigma}_{\mathbf{x}}\mathbf{A}^T\mathbf{B}^T + \boldsymbol{\Gamma}\mathbb{E}\{ \mathbf{x}\boldsymbol{\eta}^T\}\right)\left(\boldsymbol{\Sigma}_{\mathbf{z}\mathbf{x}}\mathbf{A}^T\right)^{-1}\nonumber\\
    =& \boldsymbol{\Gamma}\boldsymbol{\Sigma}_{\mathbf{x}}\mathbf{A}^T\mathbf{B}^T(\mathbf{A}\boldsymbol{\Sigma}_{\mathbf{x}}\mathbf{A}^T)^{-1}\mathbf{B}^{-1}\nonumber\\
    =& \boldsymbol{\Gamma}\boldsymbol{\Sigma}_{\mathbf{x}}\mathbf{A}^T(\mathbf{A}\boldsymbol{\Sigma}_{\mathbf{x}}\mathbf{A}^T)^{-1}.
\end{align}
The first line follows from the definition of a linear MMSE estimator. The second line follows from the generalized Bussgang decomposition. The third line follows from the linear task assumption $\tilde{\boldsymbol{s}} = \boldsymbol{\Gamma}\mathbf{x}$ and the relationship between $\boldsymbol{\Sigma}_{\mathbf{z}}$ and $\boldsymbol{\Sigma}_{\mathbf{z}\mathbf{x}}$ established in equation \eqref{eq:covariance_Z}. The fourth line is obtained from the Bussgang gain matrix expression in \eqref{eq:Bussgang_gain} and fact that $\mathbb{E}\{\mathbf{x}\boldsymbol{\eta}^T\} = \mathbf{0}$. To see this, we expand $\mathbb{E}\{\mathbf{x}\boldsymbol{\eta}^T\}$ as follows:
\begin{align*}
  \mathbb{E}\{\mathbf{x}\boldsymbol{\eta}^T\} = &  \mathbb{E}\{\mathbf{x}\left(\mathbf{z} - \mathbf{B}\mathbf{A}\mathbf{x}\right)^T\} \\
  =& \boldsymbol{\Sigma}_{\mathbf{zx}}^T - \boldsymbol{\Sigma}_{\mathbf{x}}\mathbf{A}^T\mathbf{B}^T\\
  =& \mathbf{0},
\end{align*}
where the third line holds because of \eqref{eq:Bussgang_gain}. Finally, the last line in \eqref{eq:B_opt_derivation} follows from the fact that $\mathbf{B}$ and $\mathbf{A}\boldsymbol{\Sigma}_{\mathbf{x}}\mathbf{A}^T$ are diagonal matrices. As such, \begin{align*}
    \mathbf{B}^T(\mathbf{A}\boldsymbol{\Sigma}_{\mathbf{x}}\mathbf{A}^T)^{-1}\mathbf{B}^{-1} =& \mathbf{B}^T\mathbf{B}^{-1}(\mathbf{A}\boldsymbol{\Sigma}_{\mathbf{x}}\mathbf{A}^T)^{-1} \\
    =& (\mathbf{A}\boldsymbol{\Sigma}_{\mathbf{x}}\mathbf{A}^T)^{-1}.
\end{align*}
Consequently, the quantizer-dependent MSE term of the linear MMSE estimator becomes
\begin{align*}
    \mathbb{E}\{||\tilde{\mathbf{s}} - \hat{\mathbf{s}}||^2\} = &  \mathbb{E}\{||\boldsymbol{\Gamma}\mathbf{x} - \mathbf{D}^\circ\mathbf{z}||^2\} \\
    =& \mathrm{Tr}\left(\boldsymbol{\Gamma}\boldsymbol{\Sigma}_{\mathbf{x}}\boldsymbol{\Gamma}^T\right) \\
    &- \mathrm{Tr}\left(\boldsymbol{\Gamma}\boldsymbol{\Sigma}_{\mathbf{x}}\mathbf{A}^T\mathbf{B}\left(\mathbf{A}\boldsymbol{\Sigma}_{\mathbf{x}}\mathbf{A}^T\right)^{-1}\mathbf{A}\boldsymbol{\Sigma}_{\mathbf{x}}\boldsymbol{\Gamma}^T\right),
\end{align*}
which proves the claim.

\section{Proof of Theorem \ref{theorem:A_opt}}\label{proof:A_opt}

Let $\mathbf{\tilde{A}} = \mathbf{A}\boldsymbol{\Sigma}_{\mathbf{x}}^{\frac{1}{2}}$. Then, the quantized-dependent MSE term becomes
\begin{align}
    \mathbf{E}\{||\tilde{\mathbf{s}} - \hat{\mathbf{s}}||^2\} =& \mathrm{Tr}\left(\boldsymbol{\tilde{\Gamma}}\boldsymbol{\tilde{\Gamma}}^T\right) \nonumber\\
    &- \mathrm{Tr}\Big(\boldsymbol{\tilde{\Gamma}}\mathbf{\tilde{A}}^T\mathbf{B}\left(\mathbf{\tilde{A}}\mathbf{\tilde{A}}^T\right)^{-1}\mathbf{\tilde{A}}\tilde{\mathbf{\Gamma}}^T\Big)\nonumber\\
    =& \mathrm{Tr}\left(\boldsymbol{\tilde{\Gamma}}\boldsymbol{\tilde{\Gamma}}^T\right) \nonumber\\
    &- \mathrm{Tr}\Big(\mathbf{\tilde{A}}\tilde{\mathbf{\Gamma}}^T\boldsymbol{\tilde{\Gamma}}\mathbf{\tilde{A}}^T\mathbf{B}\left(\mathbf{\tilde{A}}\mathbf{\tilde{A}}^T\right)^{-1}\Big),
\end{align}
where the second line comes from the cyclic property of the trace function. Since the first term is independent of $\mathbf{\tilde{A}}$, the optimization problem simplifies to
\begin{align}
    \mathbf{\tilde{A}}^\circ = \underset{\mathbf{\tilde{A}}}{\arg\max} \mathrm{Tr}\Big(\mathbf{\tilde{A}}\tilde{\mathbf{\Gamma}}^T\boldsymbol{\tilde{\Gamma}}\mathbf{\tilde{A}}^T\mathbf{B}\left(\mathbf{\tilde{A}}\mathbf{\tilde{A}}^T\right)^{-1}\Big).
\end{align}

Due to Restriction 1, $\mathbf{\tilde{A}}\mathbf{\tilde{A}}^T$ is a diagonal matrix. Thus, we can also represent $\mathbf{\tilde{A}}$ as
\begin{align}
    \mathbf{\tilde{A}} = \mathbf{H}_{\mathbf{\tilde{A}}}\mathbf{V}^T,
\end{align}
where $\mathbf{H}_{\mathbf{\tilde{A}}}\in \mathbb{R}^{P\times N}$ is a scaling matrix whose off-diagonal entries are zero and whose $p$-th entry in the main diagonal corresponds to a scaling of the $p$-th output of the analog combining matrix. The matrix $\mathbf{V}\in \mathbb{R}^{N\times N}$ is a unitary matrix. Under this setting, we can reduce the optimization problem to
\begin{align*}
    &\mathbf{H}^\circ,\mathbf{V}^\circ= \underset{\mathbf{H}_{\mathbf{\tilde{A}}},\mathbf{V}}{\arg\max}\;\;\mathrm{Tr}\Big(\mathbf{H}_{\mathbf{\tilde{A}}}\mathbf{V}^T\tilde{\mathbf{\Gamma}}^T\boldsymbol{\tilde{\Gamma}}\mathbf{V}\mathbf{H}_{\mathbf{\tilde{A}}}^T\mathbf{B}\left(\mathbf{H}_{\mathbf{\tilde{A}}}\mathbf{H}_{\mathbf{\tilde{A}}}^T\right)^{-1}\Big).
\end{align*}

Due to \cite[Theorem II.1]{Lasserre:1995}, $\mathbf{V}^\circ$ is the matrix containing the right singular vectors of $\tilde{\mathbf{\boldsymbol{\Gamma}}}$. This further simplifies the optimization problem to
\begin{align*}
    \mathbf{H}^\circ=  \underset{\mathbf{H}_{\mathbf{\tilde{A}}}}{\arg\max}\;\;\sum_{i = 1}^{\min(P,K)}\lambda_{\boldsymbol{\tilde{\Gamma}},i}[1-\rho_{\mathrm{q}}^{(i)}],
\end{align*}
where $\lambda_{\boldsymbol{\tilde{\Gamma}},i}$ is the $i$-th eigenvalue of $\boldsymbol{\tilde{\Gamma}}^T\boldsymbol{\tilde{\Gamma}}$ (arranged in descending order). It can be observed that the new objective function is independent of $\mathbf{H}_{\mathbf{\tilde{A}}}$ as long as the entries of main diagonal of $\mathbf{H}_{\mathbf{\tilde{A}}}$ are positive (otherwise, $\mathbf{H}_{\mathbf{\tilde{A}}}\mathbf{H}_{\mathbf{\tilde{A}}}^T$ will not be invertible). Without loss of generality, we set $\mathbf{H}^\circ = [\mathbf{I}_{P\times P}\; \mathbf{0}_{P\times (N - P)}]$. Consequently, we get
\begin{align*}
    \mathbf{A}^\circ =& \mathbf{H}^\circ(\mathbf{V}^\circ)^T\boldsymbol{\Sigma}_{\mathbf{x}}^{-\frac{1}{2}}\\
    =& \mathbf{V}_{\mathrm{opt}}^T\boldsymbol{\Sigma}_{\mathbf{x}}^{-\frac{1}{2}},
\end{align*}
where the rows of $\mathbf{V}_{\mathrm{opt}}^T$ are the $P$ right singular vectors of $\boldsymbol{\tilde{\Gamma}} = \boldsymbol{\Gamma}\boldsymbol{\Sigma}_{\mathbf{x}}^{\frac{1}{2}}$ corresponding to the $P$ largest singular values. To verify that $\mathbf{A}^\circ$ satisfies Restriction \ref{assumption:analogcombiner}, note that
\begin{align*}
    \mathbf{A}^\circ \boldsymbol{\Sigma}_{\mathbf{x}}(\mathbf{A}^\circ)^T =& \mathbf{V}_{\mathrm{opt}}^T\boldsymbol{\Sigma}_{\mathbf{x}}^{-\frac{1}{2}}\boldsymbol{\Sigma}_{\mathbf{x}}\boldsymbol{\Sigma}_{\mathbf{x}}^{-\frac{1}{2}}\mathbf{V}_{\mathrm{opt}}\\
    =& \mathbf{V}_{\mathrm{opt}}^T\mathbf{V}_{\mathrm{opt}}.
\end{align*}
Since the singular vectors are orthogonal to each other, then $\mathbf{V}_{\mathrm{opt}}^T\mathbf{V}_{\mathrm{opt}}$ is a diagonal matrix.

By plugging in $\mathbf{A}^\circ$ to $\mathbf{D}^\circ(\mathbf{A})$, we get
\begin{align}
  \mathbf{D}^\circ(\mathbf{A}^\circ) = \boldsymbol{\Gamma}\boldsymbol{\Sigma}_{\mathbf{x}}^{\frac{1}{2}}\mathbf{V}_{\mathrm{opt}}.
\end{align}

Finally, the quantizer dependent MSE can be written as
\begin{align*}
    &\mathbb{E}\{||\tilde{\mathbf{s}} - \hat{\mathbf{s}}||^2\} \\
    &\qquad= \sum_{i = 1}^{K}\lambda_{\boldsymbol{\tilde{\Gamma}},i} - \sum_{i = 1}^{\min(K,P)}\lambda_{\boldsymbol{\tilde{\Gamma}},i}\cdot[1-\rho_{\mathrm{q}}^{(i)}]\\
    &\qquad= \begin{cases}\sum_{i = 1}^{K}\lambda_{\boldsymbol{\tilde{\Gamma}},i}\cdot\rho_{\mathrm{q}}^{(i)},\quad\qquad\qquad\quad,\mathrm{if }\;P \geq K\\
    \sum_{i = 1}^{P}\lambda_{\boldsymbol{\tilde{\Gamma}},i}\cdot\rho_{\mathrm{q}}^{(i)} + \sum_{i = P+1}^{K}\lambda_{\boldsymbol{\tilde{\Gamma}},i},\;\mathrm{otherwise}.
    \end{cases}
\end{align*}

%It can be observed that the objective function is independent of $\mathbf{H}_{\mathbf{\tilde{A}}}$ as long as it assigns non-zero scaling to all columns of $\mathbf{V}_{\mathrm{opt}}$ (otherwise, $\mathbf{H}_{\mathbf{\tilde{A}}}\mathbf{H}_{\mathbf{\tilde{A}}}^T$ will not be invertible). Without loss of generality, we set $\mathbf{H}_{\mathbf{\tilde{A}}} = [\mathbf{I}_{P\times P}\; \mathbf{0}_{P\times (N - P)}]$.

%, and $R_{\mathbf{H}_{\mathbf{\tilde{A}}}} = \mathrm{rank}(\mathbf{H}_{\mathbf{\tilde{A}}}\mathbf{H}_{\mathbf{\tilde{A}}}^T)$. It can be observed that the objective function is independent of $\mathbf{H}_{\mathbf{\tilde{A}}}$

%It can be observed that the objective function is independent of $\mathbf{H}_{\mathbf{\tilde{A}}}$ as long as it assigns non-zero scaling to all columns of $\mathbf{V}_{\mathrm{opt}}$ (otherwise, $\mathbf{H}_{\mathbf{\tilde{A}}}\mathbf{H}_{\mathbf{\tilde{A}}}^T$ will not be invertible).

%Let $\mathbf{\tilde{A}} = \mathbf{C}_{\mathbf{A}\mathbf{x}}\mathbf{V}^T$, where $\mathbf{C}_{\mathbf{A}\mathbf{x}} \in \mathbb{R}^{P\times N}$ is a scaling matrix whose off-diagonal entries are zero and the $p$-th entry in the main diagonal is
% \begin{align*}
%   \mathbf{C}_{\mathbf{A}\mathbf{x}}^{(p)} = 
% \end{align*}
% $\mathbf{V}\in \mathbb{R}^{N\times N}$ is a unitary matrix.

\end{appendices}

\ifCLASSOPTIONcaptionsoff
  \newpage
\fi

\bibliographystyle{ieeetr}
\bibliography{references}

\end{document}